\documentclass[10pt,journal,compsoc]{IEEEtran}

\usepackage{graphicx}
\usepackage{tikz}
\usetikzlibrary{arrows.meta, positioning, calc, backgrounds, patterns, fit}
\usetikzlibrary{decorations}
\usetikzlibrary{decorations.pathmorphing}
\usepackage[T1]{fontenc}
\usepackage[utf8]{inputenc}

\usepackage{subcaption}
\usepackage{xcolor}
\usepackage{hyperref}

\hypersetup{
    colorlinks=true,
    citecolor=green,
    linkcolor=blue,
    urlcolor=blue,
    pdfborder={0 0 0}
}

\usepackage{setspace}
\usepackage{algorithm}
\usepackage{algpseudocode}
\usepackage{amsmath}
\usepackage{amssymb}
\usepackage{booktabs}
\usepackage{cite}  % Ensures IEEE-style citations work properly

\definecolor{myBoxColor}{HTML}{FF7F50}   % Coral
\definecolor{myProcColor}{HTML}{8FBC8F}  % DarkSeaGreen
\definecolor{myTextColor}{HTML}{2F4F4F}  % DarkSlateGray

\title{What Every Computer Scientist Needs to Know About Parallelization}

\author{%
    Temitayo~Adefemi%
    \IEEEcompsocitemizethanks{\IEEEcompsocthanksitem T.~Adefemi is with the University of Edinburgh. \protect\\
    E-mail: \texttt{T.M.Adefemi@sms.ed.ac.uk}}% 
}% end author

\begin{document}

\IEEEtitleabstractindextext{%
    \begin{abstract}
    Parallelization has become a cornerstone of modern computing, influencing everything from high-performance supercomputers to everyday mobile devices. This paper presents a comprehensive guide on the fundamentals of parallelization that every computer scientist should know, beginning with a historical perspective that traces the evolution from early theoretical models such as PRAM and BSP to today’s advanced multicore and heterogeneous architectures. We explore essential theoretical frameworks, practical paradigms, and synchronization mechanisms while discussing implementation strategies using processes, threads, and modern models like the Actor framework. Additionally, we examine how hardware components—including CPUs, caches, memory, and accelerators interact with software to impact performance, scalability, and load balancing. This work demystifies parallel programming by integrating historical context, theoretical underpinnings, and practical case studies. It equips readers with the tools to design, optimize, and troubleshoot parallel applications in an increasingly concurrent computing landscape.
    \end{abstract}

    % *** INDEX TERMS ***
}

\maketitle
\thispagestyle{plain}

\section{Introduction}
Since IBM's 1958 research memo, parallel programming has been fundamental to computing, yet its perception has often been skewed. Many computer scientists have historically viewed parallelization as an advanced, specialized concept reserved for high-performance computing (HPC) environments \cite{chandy1988programming}. This could not be further from the truth. Today, parallelization is so deeply embedded in modern computing that nearly every system relies on it in some form. From supercomputers to smartphones, cloud platforms to everyday appliances, parallel processing is everywhere \cite{mattson2004patterns}.

\vspace{0.3cm}

Despite its ubiquity, parallel programming remains underutilized by many developers, often because of the misconception that it requires esoteric knowledge or is only relevant for those working on massive-scale computations. In reality, mastering parallelization is crucial—not just for optimizing HPC workloads but for improving efficiency across all computing domains, including software development, machine learning, and embedded systems \cite{sen2012developing}.

\vspace{0.3cm}

Breaking down the barriers to understanding parallel computing is crucial to bridge this gap. This paper aims to demystify parallel computing, providing a comprehensive understanding of its principles and applications. We will explore the key factors influencing parallel programs, including parallel paradigms, hardware considerations, memory hierarchies, and cache behavior \cite{gallopoulos2016parallel}. We will also examine the trade-offs involved, the problems that benefit from parallelization, and the libraries and tools that make it accessible to developers \cite{fields2022introduction}. By the end of this paper, readers will not only grasp the abstract concepts governing parallel computing but also gain the practical knowledge to implement efficient, scalable parallel programs.

\vspace{0.3cm}

\section{What is Parallel Computing}

It is crucial to explain parallel computing first to understand it. In serial computing, a problem is broken into a discrete series of instructions executed sequentially—one after another—on a single processor. Only one instruction may be executed at any given moment, meaning tasks are processed strictly linearly. While this model is simple and effective for many applications, it can become inefficient when dealing with large-scale computational problems that require significant processing power \cite{rastogi2016significance}.

\vspace{0.4cm}

\begin{figure}[ht]
    \centering
    \begin{tikzpicture}[
        scale=0.6,
        transform shape,
        font=\small,
        node distance=1.0cm,
        >=Stealth,
        align=center,
        every node/.style={
            draw,
            rectangle,
            minimum height=0.8cm,
            minimum width=1.6cm
        }
    ]
        %--- Top Diagram: Generic problem -> instructions -> processor
        
        % "problem" block
        \node[fill=gray!20] (problem) {problem};

        % instruction blocks
        \node[right=1.2cm of problem] (tN) {t\textsubscript{N}};
        \node[right=0.8cm of tN] (t3) {t3};
        \node[right=0.8cm of t3] (t2) {t2};
        \node[right=0.8cm of t2] (t1) {t1};

        % "processor" block in orange
        \node[fill=orange!30, right=1.2cm of t1] (processor) {processor};

        % Arrows for top diagram
        \draw[->, thick] (problem) -- (tN);
        \draw[->, thick] (tN) -- (t3);
        \draw[->, thick] (t3) -- (t2);
        \draw[->, thick] (t2) -- (t1);
        \draw[->, thick] (t1) -- (processor);
        
        %--- Bottom Diagram: Example with process_data()
        
        % Place the example label well below the top diagram
        \node[draw=none, below=2.5cm of problem, xshift=2.3cm] (exampleLabel)
            {\textbf{For example:} \texttt{process\_data()}};

        % instruction blocks for example
        \node[below=1.0cm of exampleLabel] (tn_ex) {t\textsubscript{N}\_load};
        \node[right=0.8cm of tn_ex] (t3_ex) {t3\_filter};
        \node[right=0.8cm of t3_ex] (t2_ex) {t2\_aggregate};
        \node[right=0.8cm of t2_ex] (t1_ex) {t1\_store};

        % processor block in orange
        \node[fill=orange!30, right=1.2cm of t1_ex] (processor_ex) {processor};

        % Arrows for bottom diagram
        % Using a simple arrow from the example label to the first instruction
        \draw[->, thick] (exampleLabel) -- (tn_ex);
        \draw[->, thick] (tn_ex) -- (t3_ex);
        \draw[->, thick] (t3_ex) -- (t2_ex);
        \draw[->, thick] (t2_ex) -- (t1_ex);
        \draw[->, thick] (t1_ex) -- (processor_ex);

    \end{tikzpicture}
    \caption{A schematic showing how a problem is broken down into instructions and executed by a processor. An example function \texttt{process\_data()} is also shown.}
    \label{fig:instructions_processor}
\end{figure}
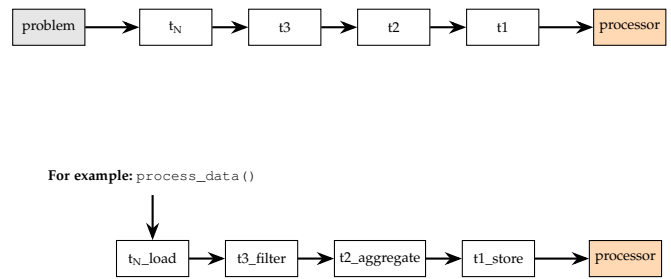

Parallel computing, in contrast, is the simultaneous use of multiple computing resources to solve a computational problem. The problem is divided into discrete parts that can be solved concurrently, and each part is further decomposed into a series of instructions. These instructions are executed simultaneously across different processors, leveraging the computational power of multi-core CPUs, GPUs, or distributed computing systems \cite{mallegowda2024efficiency}. An overall control/coordination mechanism maintains accuracy and synchronization, ensuring that each task progresses as intended. However, the complexity of these mechanisms varies based on the problem being solved and the parallelization paradigm used—ranging from shared memory models with synchronization primitives to distributed computing models that use message passing for inter-process communication \cite{wu2012parallelizing}.

\vspace{0.3cm}

\begin{figure}[ht]
\centering
\begin{tikzpicture}[
    font=\small,
    node distance=1.0cm,
    >=Stealth,
    thick,
    align=center,
    scale=0.7,
    every node/.style={transform shape}
]

%%%%%%%%%%%%%%%%%%%%%%%%%%%%%%%%%%%%%%%%%%%%%%%%%%%%%%%%%%%%
% COLUMN REFERENCES
%%%%%%%%%%%%%%%%%%%%%%%%%%%%%%%%%%%%%%%%%%%%%%%%%%%%%%%%%%%%
% Defining column positions for spacing consistency
\coordinate (colProblem) at (0,0);     
\coordinate (colInstrStart) at (4,0);  
\coordinate (colProcessor) at (10,0);  

%%%%%%%%%%%%%%%%%%%%%%%%%%%%%%%%%%%%%%%%%%%%%%%%%%%%%%%%%%%%
% ROW COORDINATES FOR SPACING
%%%%%%%%%%%%%%%%%%%%%%%%%%%%%%%%%%%%%%%%%%%%%%%%%%%%%%%%%%%%
\coordinate (row1) at (0,-1.0);
\coordinate (row2) at (0,-2.8);
\coordinate (row3) at (0,-4.6);
\coordinate (row4) at (0,-6.4);

%%%%%%%%%%%%%%%%%%%%%%%%%%%%%%%%%%%%%%%%%%%%%%%%%%%%%%%%%%%%
% ROW 1: process_data(emp1)
%%%%%%%%%%%%%%%%%%%%%%%%%%%%%%%%%%%%%%%%%%%%%%%%%%%%%%%%%%%%
\node[rectangle, draw=black, fill=white,
      minimum width=3cm, minimum height=1cm,
      anchor=west] (prob1)
      at (colProblem |- row1)
      {\texttt{process\_data(emp1)}};

% Instructions
\node[rectangle, draw=black, fill=white,
      minimum width=1cm, minimum height=1cm,
      anchor=west] (tN1)
      at (colInstrStart |- row1) {tN};

\node[rectangle, draw=black, fill=white,
      minimum width=1cm, minimum height=1cm,
      anchor=west, xshift=0.3cm] (t31)
      at (tN1.east) {t3};

\node[rectangle, draw=black, fill=white,
      minimum width=1cm, minimum height=1cm,
      anchor=west, xshift=0.3cm] (t21)
      at (t31.east) {t2};

\node[rectangle, draw=black, fill=white,
      minimum width=1cm, minimum height=1cm,
      anchor=west, xshift=0.3cm] (t11)
      at (t21.east) {t1};

% Processor (light blue for variety)
\node[rectangle, draw=black, fill=orange!30,
      minimum width=2cm, minimum height=1cm,
      anchor=west] (proc1)
      at (colProcessor |- row1) {processor};

% Arrows
\draw[->] (prob1.east) -- (tN1.west);
\draw[->] (tN1.east) -- (t31.west);
\draw[->] (t31.east) -- (t21.west);
\draw[->] (t21.east) -- (t11.west);
\draw[->] (t11.east) -- (proc1.west);

%%%%%%%%%%%%%%%%%%%%%%%%%%%%%%%%%%%%%%%%%%%%%%%%%%%%%%%%%%%%
% ROW 2: process_data(emp2)
%%%%%%%%%%%%%%%%%%%%%%%%%%%%%%%%%%%%%%%%%%%%%%%%%%%%%%%%%%%%
\node[rectangle, draw=black, fill=white,
      minimum width=3cm, minimum height=1cm,
      anchor=west] (prob2)
      at (colProblem |- row2)
      {\texttt{process\_data(emp2)}};

\node[rectangle, draw=black, fill=white,
      minimum width=1cm, minimum height=1cm,
      anchor=west] (tN2)
      at (colInstrStart |- row2) {tN};

\node[rectangle, draw=black, fill=white,
      minimum width=1cm, minimum height=1cm,
      anchor=west, xshift=0.3cm] (t32)
      at (tN2.east) {t3};

\node[rectangle, draw=black, fill=white,
      minimum width=1cm, minimum height=1cm,
      anchor=west, xshift=0.3cm] (t22)
      at (t32.east) {t2};

\node[rectangle, draw=black, fill=white,
      minimum width=1cm, minimum height=1cm,
      anchor=west, xshift=0.3cm] (t12)
      at (t22.east) {t1};

\node[rectangle, draw=black, fill=orange!30,
      minimum width=2cm, minimum height=1cm,
      anchor=west] (proc2)
      at (colProcessor |- row2) {processor};

\draw[->] (prob2.east) -- (tN2.west);
\draw[->] (tN2.east) -- (t32.west);
\draw[->] (t32.east) -- (t22.west);
\draw[->] (t22.east) -- (t12.west);
\draw[->] (t12.east) -- (proc2.west);

%%%%%%%%%%%%%%%%%%%%%%%%%%%%%%%%%%%%%%%%%%%%%%%%%%%%%%%%%%%%
% ROW 3: process_data(emp3)
%%%%%%%%%%%%%%%%%%%%%%%%%%%%%%%%%%%%%%%%%%%%%%%%%%%%%%%%%%%%
\node[rectangle, draw=black, fill=white,
      minimum width=3cm, minimum height=1cm,
      anchor=west] (prob3)
      at (colProblem |- row3)
      {\texttt{process\_data(emp3)}};

\node[rectangle, draw=black, fill=white,
      minimum width=1cm, minimum height=1cm,
      anchor=west] (tN3)
      at (colInstrStart |- row3) {tN};

\node[rectangle, draw=black, fill=white,
      minimum width=1cm, minimum height=1cm,
      anchor=west, xshift=0.3cm] (t33)
      at (tN3.east) {t3};

\node[rectangle, draw=black, fill=white,
      minimum width=1cm, minimum height=1cm,
      anchor=west, xshift=0.3cm] (t23)
      at (t33.east) {t2};

\node[rectangle, draw=black, fill=white,
      minimum width=1cm, minimum height=1cm,
      anchor=west, xshift=0.3cm] (t13)
      at (t23.east) {t1};

\node[rectangle, draw=black, fill=orange!30,
      minimum width=2cm, minimum height=1cm,
      anchor=west] (proc3)
      at (colProcessor |- row3) {processor};

\draw[->] (prob3.east) -- (tN3.west);
\draw[->] (tN3.east) -- (t33.west);
\draw[->] (t33.east) -- (t23.west);
\draw[->] (t23.east) -- (t13.west);
\draw[->] (t13.east) -- (proc3.west);

%%%%%%%%%%%%%%%%%%%%%%%%%%%%%%%%%%%%%%%%%%%%%%%%%%%%%%%%%%%%
% ROW 4: process_data(empN)
%%%%%%%%%%%%%%%%%%%%%%%%%%%%%%%%%%%%%%%%%%%%%%%%%%%%%%%%%%%%
\node[rectangle, draw=black, fill=white,
      minimum width=3cm, minimum height=1cm,
      anchor=west] (prob4)
      at (colProblem |- row4)
      {\texttt{process\_data(empN)}};

\node[rectangle, draw=black, fill=white,
      minimum width=1cm, minimum height=1cm,
      anchor=west] (tN4)
      at (colInstrStart |- row4) {tN};

\node[rectangle, draw=black, fill=white,
      minimum width=1cm, minimum height=1cm,
      anchor=west, xshift=0.3cm] (t34)
      at (tN4.east) {t3};

\node[rectangle, draw=black, fill=white,
      minimum width=1cm, minimum height=1cm,
      anchor=west, xshift=0.3cm] (t24)
      at (t34.east) {t2};

\node[rectangle, draw=black, fill=white,
      minimum width=1cm, minimum height=1cm,
      anchor=west, xshift=0.3cm] (t14)
      at (t24.east) {t1};

\node[rectangle, draw=black, fill=orange!30,
      minimum width=2cm, minimum height=1cm,
      anchor=west] (proc4)
      at (colProcessor |- row4) {processor};

\draw[->] (prob4.east) -- (tN4.west);
\draw[->] (tN4.east) -- (t34.west);
\draw[->] (t34.east) -- (t24.west);
\draw[->] (t24.east) -- (t14.west);
\draw[->] (t14.east) -- (proc4.west);

\end{tikzpicture}
\caption{Parallel execution of \texttt{process\_data()} for multiple inputs. Each input (e.g., \texttt{emp1}) is processed separately in parallel.}
\end{figure}
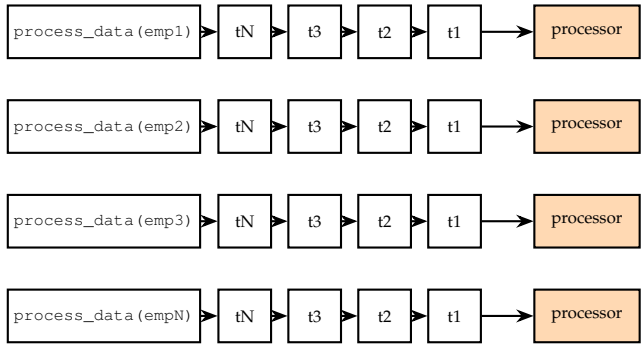

\vspace{0.3cm}

Based on this definition, it becomes evident why a coordination mechanism is essential to ensure the program behaves as intended. These mechanisms operate at multiple levels of the parallel computing stack, from software-level parallel programming models to hardware-level synchronization protocols. At the lowest level, threads use locks, mutexes, and semaphores to prevent simultaneous updates to shared resources, which could lead to race conditions and data inconsistencies \cite{jian2002parallel}. Higher up the stack, message-passing mechanisms—such as those implemented in MPI (Message Passing Interface)—allow distributed processes to communicate effectively and synchronize their execution \cite{zhou2005numerical}.

\vspace{0.3cm}

Why parallelize a program? The primary motivation is to enhance speed and efficiency. Modern computers typically feature multi-core processors, GPUs, and high-performance distributed systems, making it logical to share computational loads across multiple processing units rather than allowing resources to remain idle. By distributing workloads efficiently, programs can significantly reduce execution times, achieve higher throughput, and better use available hardware resources \cite{rathod2014serial}.

\section{Historical Perspective}

In order to give an overview of parallelization, it is important to understand its roots. Parallelization can be traced back to the early days of computing in the 1950s when researchers first recognized the need to process multiple tasks concurrently, which brought forward the IBM research memo, as discussed earlier. The early pioneers began exploring methods to execute several operations simultaneously, which set the stage for modern parallel computing \cite{nikolic2022fifty}.

\vspace{0.3cm}

As computer architectures advanced in the 1960s and 1970s, so did the ideas surrounding parallel computation. Researchers developed early models such as the Parallel Random Access Machine (PRAM), which provided a simplified abstraction for understanding how multiple processors could work together on a single problem. However, PRAM had its limitations as it did not consider the complexities surrounding parallel computing. During this era, large mainframe systems and specialized supercomputers were built with parallel processing capabilities, albeit within very controlled environments, paving the way for more ambitious applications in scientific and engineering domains \cite{gepner2006multi}.

\vspace{0.3cm}

The 1980s witnessed significant progress in hardware design and algorithm development, leading to the advent of dedicated parallel machines. This period saw the emergence of vector processors and early multiprocessor systems that could handle more complex, data-intensive tasks. Academic research and government-funded projects contributed to a deeper understanding of synchronization, load balancing, and the challenges of distributed memory, which are still crucial to parallel computing today \cite{dangelo2014new}.

\vspace{0.3cm}

With the advent of microprocessors in the 1990s, parallelization transitioned from specialized supercomputers to more widely available commodity hardware. The introduction of multi-core processors revolutionized the computing landscape, making parallel processing accessible to a broader audience. This shift was accompanied by the development of robust programming models and standards, such as MPI and OpenMP, which allowed developers to exploit parallelism more easily in everyday applications \cite{chennupati2014multi}.

\vspace{0.3cm}

Today, parallelization is a fundamental aspect of nearly every computing system, from high-performance clusters to smartphones. The historical evolution from theoretical models and expensive hardware to ubiquitous, multi-core devices underscores the transformative impact of parallel computing. Modern computer scientists benefit from decades of research and practical advancements that have made parallel programming a specialized skill and an essential component of practical software development across diverse fields \cite{ghose2024general}.

\section{Theories Governing Parallel Computing}

\subsection{The PRAM Model}

One of the earliest and most influential theoretical frameworks in parallel computing is the \textit{Parallel Random Access Machine (PRAM)} model, which was mentioned in the historical perspective section. PRAM provides an idealized abstraction of parallel computation, where multiple processors operate synchronously and share a standard memory. The simplicity of PRAM in capturing parallelism has made it a widely used model for designing parallel algorithms \cite{hall2011pram}. However, this abstraction also introduces challenges, as full synchronization and shared memory access are costly in practical implementations \cite{karp1988survey}.

\begin{itemize}
    \item \textbf{Exclusive Read Exclusive Write (EREW):} No simultaneous reading or writing of the same memory cell.
    \item \textbf{Concurrent Read Exclusive Write (CREW):} Multiple processors can read the same cell, but only one may write.
    \item \textbf{Concurrent Read Concurrent Write (CRCW):} Both reading and writing can be performed concurrently, with various rules to resolve write conflicts.
\end{itemize}

While PRAM is primarily a theoretical construct, it has motivated the development of specialized hardware and emulation techniques. Some researchers have attempted to map PRAM models onto modern many-core processors, such as Intel's Single-chip Cloud Computer (SCC), to bridge the gap between theory and real-world implementations \cite{lecomber2000pram}. However, practical constraints, such as memory bandwidth limitations and synchronization overheads, limit direct PRAM implementations \cite{mulmuley1999lower}.

\vspace{0.3cm}

Despite its limitations, PRAM remains relevant in algorithm design and analysis, often as a stepping stone for developing practical parallel algorithms \cite{ullman1986parallel}.

\subsection{Bulk Synchronous Parallel (BSP) Model}

Developed by Leslie Valiant, the \textit{BSP model} introduces a more realistic abstraction that segments computation into a series of supersteps. Each superstep consists of three phases:

\begin{enumerate}
    \item \textbf{Local Computation:} Processors perform computations using local data.
    \item \textbf{Communication:} Data is exchanged between processors.
    \item \textbf{Barrier Synchronization:} Processors wait until all have completed the current superstep before proceeding.
\end{enumerate}

BSP explicitly captures the cost of communication and synchronization, making it a valuable tool for predicting and optimizing performance in real-world parallel systems \cite{mccoll1994bsp}.

\subsection{The LogP Model}

To further refine our understanding of parallel execution, the \textit{LogP} model introduces four key parameters that account for realistic communication and computation costs:

\begin{itemize}
    \item \textbf{L (Latency):} The delay incurred in communicating a message from one processor to another.
    \item \textbf{o (Overhead):} The time a processor spends sending or receiving a message.
    \item \textbf{g (Gap):} The minimum time interval between consecutive message transmissions.
    \item \textbf{P (Processors):} The number of processors in the system.
\end{itemize}

LogP provides a practical framework for analyzing parallel execution costs, particularly in distributed-memory architectures. It addresses PRAM's shortcomings by considering real-world constraints such as network communication overhead and memory access latency \cite{culler1996logp}.

\subsection{Scalability and Speedup: Amdahl's Law and Gustafson's Law}

\subsubsection{a. Amdahl's Law}

Amdahl's Law is a foundational principle that quantifies the potential speedup of a parallel program. It states that if a fraction \( f \) of a task is inherently serial, the maximum speedup \( S \) achievable with \( P \) processors is limited by:

\[
S(P) = \frac{1}{f + \frac{1-f}{P}}
\]

This Law underscores a critical limitation: even if most of the computation can be parallelized, the serial portion restricts overall performance. As \( P \) approaches infinity, the speedup converges to \( \frac{1}{f} \) \cite{hall2011pram}.

\vspace{0.3cm}

This is critical for understanding and analyzing parallel programs and how parallelization can impact their potential performance.

\subsubsection{b. Gustafson's Law}

Gustafson's Law offers a more optimistic view by arguing that as we increase the problem size, the parallelizable portion of the workload grows, potentially mitigating the impact of the serial fraction. Instead of focusing on fixed problem sizes, Gustafson's perspective considers that we often tackle more significant problems, with more processors and the overall efficiency can improve \cite{karp1988survey}.

\subsection{Complexity Classes and Parallel Algorithms}

Understanding which problems can be efficiently parallelized involves concepts from computational complexity theory.

\subsubsection{a. The NC Class}

The class \textbf{NC} (Nick's Class) includes decision problems that can be solved in polylogarithmic time using a polynomial number of processors. These problems are considered "efficiently parallelizable," making them ideal candidates for parallel computation. Examples include parallel sorting algorithms, matrix multiplication, and prefix sum computations \cite{ullman1986parallel}.

\subsubsection{b. P-Complete Problems}

In contrast, \textbf{P-complete} problems are believed to be inherently sequential. While they can be solved in polynomial time, no known efficient parallel algorithm exists. It is widely conjectured that P-complete problems do not belong to NC, placing a fundamental limit on parallelizability \cite{mulmuley1999lower}.

\section{Parallel Computing Paradigms}

\subsection{Processes}

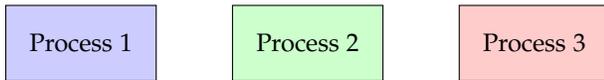
\begin{figure}[ht]
  \centering
  \begin{tikzpicture}[node distance=1cm, auto, scale=0.3]
    % Draw three colored process boxes
    \node[draw, rectangle, fill=blue!20, minimum width=2cm, minimum height=1cm] (proc1) {Process 1};
    \node[draw, rectangle, fill=green!20, minimum width=2cm, minimum height=1cm, right=of proc1] (proc2) {Process 2};
    \node[draw, rectangle, fill=red!20, minimum width=2cm, minimum height=1cm, right=of proc2] (proc3) {Process 3};
  \end{tikzpicture}
  \caption{Processes in an Operating System. Each process runs in its own isolated memory space and maintains separate resources.}
  \label{fig:processes}
\end{figure}

In computer science fundamentals, a process is a unit of execution - an 'active' entity, distinct from a program, which is a 'passive' entity. When a program (like a .exe or binary file) is run multiple times, each instance creates a new process. This concept enables parallelization, where a program can divide a problem across multiple processes. For example, consider a 100×100 2D matrix containing 10,000 elements (100 × 100). If split across 10 processes for parallel computation, each process would handle 1,000 elements \cite{hasta2010performance}.

\vspace{0.3cm}

There are multiple ways to distribute elements across processes, such as block-cyclic, block, and cyclic distribution. In block distribution, contiguous chunks of data are assigned to each process. Cyclic distribution alternates elements among processes in a round-robin fashion. Block cyclic combines these approaches by cyclically distributing blocks of elements, balancing communication overhead, and load \cite{hua2013comparison}. This hybrid approach benefits algorithms requiring local data access and regular communication patterns, such as matrix operations in parallel computing. Each of these distribution strategies typically affects the scalability and efficiency of the program. It is essential to understand the problem's core and pick the appropriate strategy based on that. It is crucial to benchmark the program with different distribution strategies when in doubt, as thorough testing is key to ensuring the best performance.

\vspace{0.3cm}

It is not only matrices that can be distributed across processes; matrices were the example given due to their prevalent use in high-performance computing. Processes are isolated, which means each unit of execution runs within its environment with no direct awareness of other processes. The only way to communicate between processes is through message passing, which ensures they can work together to solve problems. Multiple libraries implement message passing, but the most popular and standardized implementation is the Message Passing Interface (MPI). MPI provides a comprehensive set of protocols and routines that enable efficient data exchange and synchronization between parallel processes, allowing robust parallel functionality \cite{hilbrich2009mpi}.

\subsection{Threads}

\begin{figure}[ht]
  \centering
  \begin{tikzpicture}[
    node distance=1cm and 1.5cm,
    auto,
    scale=0.7
  ]
    % Draw the shared memory area for the process with a yellow background
    \node[draw, rectangle, fill=yellow!20, minimum width=4cm, minimum height=6cm] (shared) {%
      \begin{tabular}{c}
        Shared Memory\\
        (Code, Data, Heap)
      \end{tabular}
    };
    
    % Draw individual thread stacks with different colors to the right of the shared memory
    \node[draw, rectangle, fill=blue!20, minimum width=3cm, minimum height=1.2cm, 
          right=2cm of shared.north east, anchor=north west] (t1) {Thread 1 Stack};
    
    \node[draw, rectangle, fill=green!20, minimum width=3cm, minimum height=1.2cm,
          right=2cm of shared, anchor=west] (t2) {Thread 2 Stack};
    
    \node[draw, rectangle, fill=red!20, minimum width=3cm, minimum height=1.2cm,
          right=2cm of shared.south east, anchor=south west] (t3) {Thread 3 Stack};
    
    % Connect the shared memory to each thread stack with dashed arrows
    \draw[-stealth, dashed, thick] (shared.north east) -- (t1.west);
    \draw[-stealth, dashed, thick] (shared.east) -- (t2.west);
    \draw[-stealth, dashed, thick] (shared.south east) -- (t3.west);
  \end{tikzpicture}
  \caption{Threads in an Operating System. Threads share the process's memory (code, data, heap) while each maintains its own stack for execution context.}
  \label{fig:threads}
\end{figure}
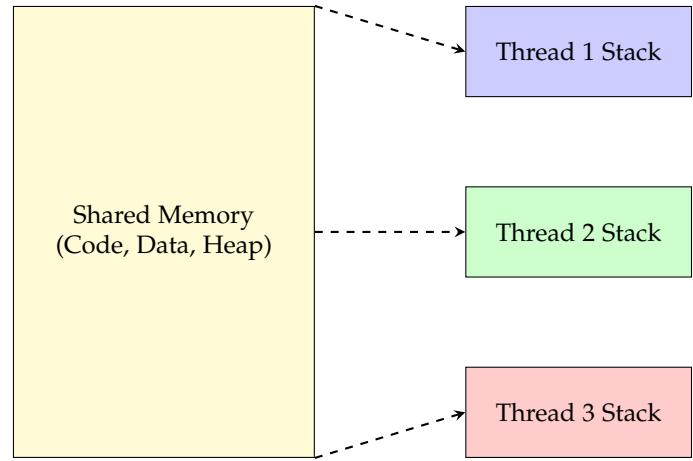
\subsection{Threads in Parallel Computing}

A thread refers to a single sequential flow of execution within a process. Threads are typically not as isolated as processes since they exist within the same memory space and share resources like heap memory, file handles, and global variables. They are more lightweight than processes because creating and switching between threads requires less overhead, but there are limits to the number of threads that can be efficiently run on a system. These limitations stem from various factors, including available system resources (memory and CPU cores), operating system constraints, and the overhead of context switching between threads \cite{jin2003multidimensional, bocian2018openmp}.

\vspace{0.3cm}

Threads can be used to solve various parallel computing problems. One example is the adaptive quadrature problem, which can be solved using a First In, First Out (FIFO) approach where threads cooperatively process work items from a shared queue until the desired accuracy threshold is reached; each thread can also have its own queue. However, this is not the only threading-based solution - the same problem can also be solved using higher-level abstractions like tasks, which provide a more structured approach to parallel execution. Modern threading frameworks and libraries have evolved to provide increasingly sophisticated abstractions and primitives, making implementing parallel algorithms efficiently while handling synchronization, load balancing, and thread safety concerns easier \cite{zhang2005thread, shirako2011phasers}.

\vspace{0.3cm}

Several libraries allow threading to occur within the operating system, and a popular standard is the OpenMP standard, which provides coordination and synchronization mechanisms for threads to solve problems accurately while working in unison. There are mechanisms such as locks that prevent race conditions by providing exclusive access to a resource to one thread at a time and other directives that help manage parallel execution, including barriers for synchronization points, critical sections for protecting shared resources, atomic operations for thread-safe updates, and scheduling clauses that control how work is distributed among threads. These features, combined with OpenMP's pragma-based approach, make it easier for developers to parallelize their code while maintaining correctness and achieving better performance \cite{maheo2012adaptive, rajput2013bottleneck}.

\section{Factors Impacting Parallelization}
Multiple factors influence the performance, efficiency, and potential for parallelizing parallel programs. This section provides an overview of these critical components and examines their extent of impact. The most crucial factor we must first consider is the nature of the problem itself—specifically, whether it can be parallelized and how well it can be parallelized \cite{foster1995designing}.

\subsection{Nature of the Problem}

The characteristics of the problem fundamentally determine whether parallelization will be effective and what approach should be taken. Some problems are inherently sequential and resist parallelization, while others naturally divide into independent tasks that can be executed concurrently \cite{quinn2004parallel}. Understanding these problem characteristics is essential before attempting to implement any parallel solution, as they directly influence the potential speedup and scalability that can be achieved through parallelization \cite{mcCool2012structured}.

\vspace{0.3cm}

The first step in parallelization is to find concurrency in the problem. Some problems are embarrassingly parallel, while others cannot be parallelized even with the most robust and innovative methods \cite{herlihy2020art}. An example of an embarrassingly parallel program is 3D video rendering handled by a graphics processing unit, where each frame (forward method) or pixel (ray tracing method) can be handled with no interdependency, allowing the workload to be distributed across multiple processors with minimal overhead \cite{owens2007survey}.

\vspace{0.3cm}

If a problem has a chain of dependencies—meaning that you must compute step $i$ before step $i+1$—then there is little opportunity to run parts of the problem simultaneously \cite{kulkarni2007optimistic}. For example, many recursive or iterative processes where the output of one iteration is needed for the next are challenging to parallelize, as the sequential nature of the algorithm forces a strict order of operations that limits concurrency \cite{liavas1998parallel}.

\vspace{0.3cm}

We have discussed the NC (Nick's Class), which consists of problems that can be solved in polylogarithmic time using a polynomial number of processors \cite{greenlaw1995limits}. P-complete problems are believed to be inherently sequential because finding an efficient parallel (NC) algorithm for any P-complete problem would imply that P equals NC, a result that most experts doubt \cite{garey1979computers}. A classic example of a P-complete problem is the Circuit Value Problem (CVP), a benchmark for problems resistant to parallel approaches \cite{greenlaw1995limits}.

\vspace{0.3cm}

There are many problems where concurrency can be found to parallelize the program; problems that do not have a sequential interdependency have massive potential for parallelization \cite{blythe2003task}. Data structures such as arrays, graphs, queues, stacks, and hash tables often support a wide range of parallel patterns, which will be discussed later in this paper \cite{cormen2022introduction}. This inherent ability to divide work among independent subtasks is what makes them attractive targets for parallel processing.

\vspace{0.3cm}

For instance, when processing large datasets, operations can often be performed on different data segments simultaneously without affecting the final result \cite{dean2008mapreduce}. Similarly, many simulation problems can be divided spatially, with different regions being computed independently before combining results \cite{foster1995designing}. This division into independent tasks allows for efficient utilization of computing resources, provided that the data can be partitioned in a balanced and effective manner \cite{quinn2004parallel}.

\vspace{0.3cm}

The challenge lies in identifying these independent components and determining the optimal granularity of parallelization, as too fine-grained parallelism can lead to excessive overhead \cite{kulkarni2007optimistic}. At the same time, parallelism that is too coarse-grained might not fully utilize available computing resources. Understanding the nature of these parallelizable problems is essential before applying specific parallel programming patterns and techniques \cite{mcCool2012structured}.  Illustrating how different data structures can be parallelized.

\subsection{Problem Size}
Another significant factor influencing the efficiency of parallelization is the size of the problem. For example, if a task operates on a single element and cannot be subdivided into smaller, independent sub-tasks, then parallel processing might offer little to no benefit. In such cases, the workload is too fine-grained to distribute effectively across multiple processing units \cite{hill2013parallel}. This issue arises because the overhead of managing parallel execution, including synchronization and inter-process communication, outweighs any potential speedup \cite{barney2010introduction}.

\vspace{0.3cm}

Conversely, many problems are inherently large-scale, presenting their challenges regarding parallelization. When a dataset or computational task is huge, simply dividing it into chunks might not be enough. Two major issues can arise:

\begin{enumerate}
\item \textbf{Memory Constraints:} After partitioning the data, each chunk must fit into the available memory. If the dataset is so vast that the divided portions exceed the memory capacity, the system may resort to disk swapping or other slower memory management techniques \cite{fu1997space}. This negates the speed benefits of parallelization and can lead to significant performance bottlenecks. Efficient execution of large-scale parallel tasks under memory constraints requires intelligent memory management techniques, such as active memory scheduling and hierarchical memory models, to maximize utilization while reducing latency \cite{song2009parallel}.

\item \textbf{Cache Efficiency:} Modern processors rely heavily on cache memory to speed up data access. Maintaining cache coherence becomes critical when a problem is divided among multiple processors or threads \cite{hennessy2011computer}. For huge problems, frequent cache invalidations and the overhead of synchronizing caches across cores can severely degrade performance \cite{hill2013parallel}. Research indicates that cache coherence protocols and optimal cache sizing significantly impact the performance of parallel algorithms, and choosing the proper configuration can lead to substantial improvements in computational efficiency \cite{fagundes2024evaluation}. Additionally, optimizing cache-aware algorithms, such as locality-preserving data structures, can help reduce cache misses and improve speed \cite{arge2008fundamental}.

\end{enumerate}
\vspace{0.3cm}

In summary, while parallelization can dramatically accelerate computational tasks, its effectiveness depends on the problem size \cite{hill2013parallel}. Tasks that are too small may not be divisible into enough sub-tasks to justify the overhead of parallel processing. On the other hand, huge problems may run into memory limitations and cache inefficiencies, both of which can limit the performance gains \cite{hennessy2011computer}. Understanding and addressing these challenges is crucial when designing systems and algorithms for practical parallel computation.

\vspace{0.3cm}

To evaluate how the size of a problem impacts parallelization, we conducted two scaling experiments on the Cirrus supercomputer at the University of Edinburgh for a cellular automaton problem that implements Conway's Game of Life parallelized using Message Passing Interface. The first experiment examined weak scaling by maintaining a fixed number of 5 processors while varying the landscape size from 500×500 to 5000×5000. The second experiment tested strong scaling by fixing the landscape size at 5000×5000 while varying the number of processors from 2 to 16. Each configuration was run 10 times to ensure reliable measurements \cite{gropp1999using}. The implementation used a decomposition strategy where the grid was divided among processes, with each process responsible for updating its local portion while managing necessary boundary communications with neighboring processes \cite{thakur2005optimization}.

\begin{figure}[h]
\centering
\includegraphics[width=0.5\textwidth]{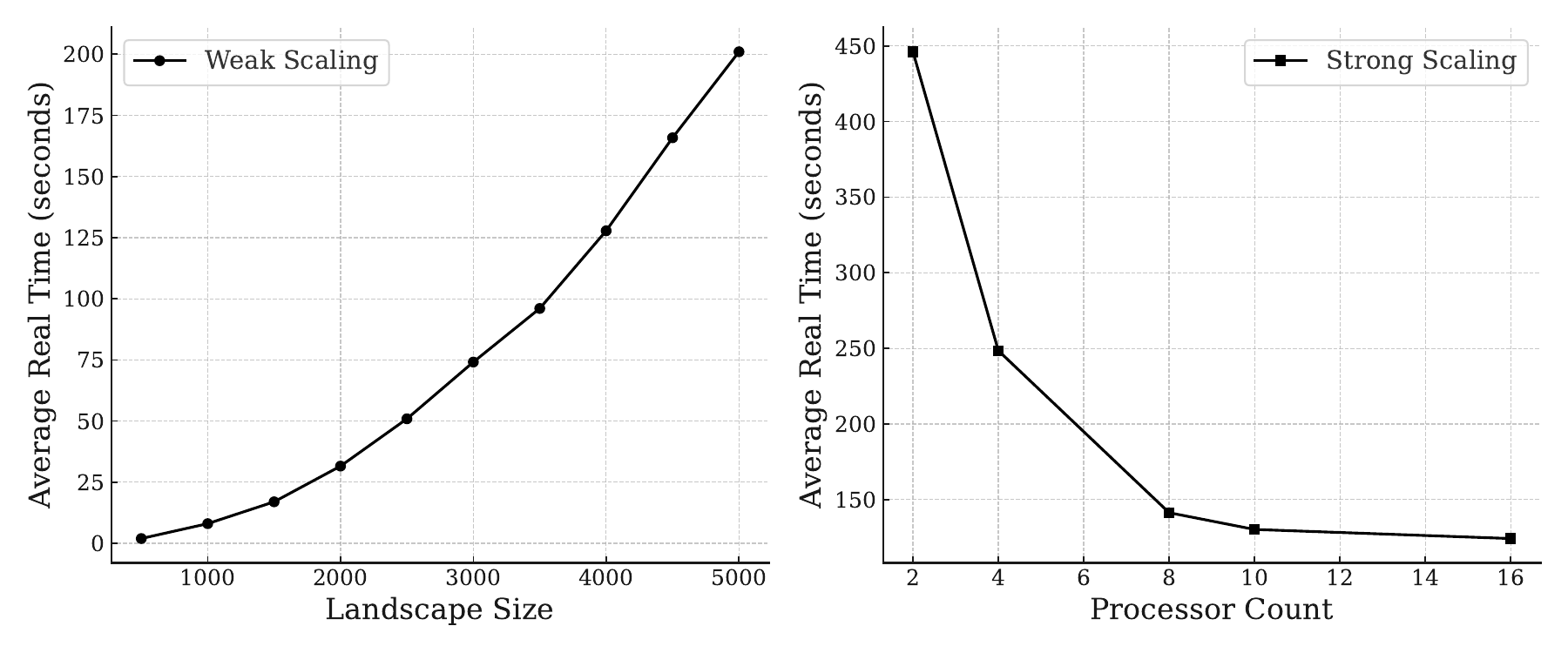}
\caption{Scaling Results of a Cellular Automaton parallelised using Message Passing Interface}
\label{fig:example}
\end{figure}

Our experiments on the Cirrus supercomputer were designed to assess how varying landscape sizes affect the parallelization of a C-based MPI implementation of Conway’s Game of Life. The simulation divides the global grid among processors, with each processor handling a subgrid that includes ghost boundaries for halo exchanges \cite{gropp1999using}. As the overall landscape size increases—from 500×500 to 5000×5000 cells—the computational load per process grows significantly, since the total number of cells (and thus the number of updates) increases quadratically with the grid dimensions \cite{quinn2004parallel}.

\vspace{0.3cm}

While larger landscapes naturally increase the volume of computations, they also affect the communication overhead inherent in MPI applications \cite{thakur2005optimization}. Each process communicates with its neighbours to update ghost cells, and the cost of these communications depends on the perimeter of the subgrid \cite{hennessy2011computer}. In larger grids, the relative ratio of the communication boundary (perimeter) to the computational workload (area) tends to decrease, potentially offering better efficiency per process \cite{hill2013parallel}. However, the absolute amount of data exchanged still grows, which can lead to longer overall runtimes despite the improved computation-to-communication ratio \cite{dean2008mapreduce}.

\vspace{0.3cm}

It is clear that increasing the landscape size intensifies the parallel workload, which typically influences the efficiency of parallel programs; this is evident, leading to a marked rise in computational effort reflected in the observed execution times \cite{gropp1999using}. Although larger local subgrids can mitigate the relative impact of communication overhead, growth in the number of cells ultimately results in increased runtimes \cite{quinn2004parallel}. This analysis underscores the need to balance computational workload and communication efficiency when scaling parallel MPI applications for larger problem sizes \cite{thakur2005optimization}. The pseudocode of the program is available in the Appendix.

\vspace{0.3cm}

\subsection{Parallel Patterns}
The parallelization pattern typically affects the parallel program's scalability, speedup, load balancing, and overhead. It is important to state that there are constructs of parallel patterns and parallel patterns themselves. There are many parallel patterns, some popular ones including geometric decomposition, actor pattern, pipeline pattern and recursive data pattern. Selecting the appropriate pattern dictates how efficiently a program scales with additional processors and influences the overall execution speed by managing overhead and balancing loads across computing units \cite{huda2021patterns}. For instance, geometric decomposition divides a problem's domain into smaller regions that can be solved concurrently, often improving scalability and speedup \cite{michalakes1999data}. At the same time, the actor pattern emphasizes independent entities communicating via messages to maintain balanced work distribution \cite{rinaldi2019actor}. Similarly, the pipeline pattern organizes tasks into a sequence of processing stages that can operate in parallel, and the recursive data pattern leverages divide-and-conquer strategies to handle complex tasks by breaking them into simpler subproblems \cite{aguilar2017recursion}. Each pattern relies on underlying constructs provided by modern programming languages and frameworks, reinforcing that low-level tools and high-level design patterns are essential for developing efficient and robust parallel programs.

\vspace{0.3cm}
Effective load balancing is another critical factor influenced by the chosen pattern. Techniques such as the Actor model encapsulate state and behavior into independent units communicating through message passing. This results in a more dynamic distribution of tasks across processors and reduces the risk of some cores idling while others are overburdened \cite{rinaldi2019actor}. However, every pattern brings its overheads; for instance, the pipeline pattern, which organizes computation into sequential stages, may suffer from inefficiencies if one stage processes data slower than the others, causing subsequent stages to wait \cite{navarro2009pipeline}.

\vspace{0.3cm}

It is also essential to distinguish between the constructs of parallel patterns and the patterns themselves. Constructs refer to the fundamental building blocks provided by programming languages or libraries—such as threads, tasks, futures, or message-passing mechanisms—that enable parallel execution \cite{collins2011datadriven}. In contrast, parallel patterns are higher-level, reusable solutions that encapsulate best practices and design strategies for common parallel programming challenges. By leveraging these patterns, developers can create efficient parallel programs without reinventing the wheel for each new problem \cite{huda2021patterns}.

\subsection{Programming Language \& Libraries}
Each programming language is suitable for specific applications. Programming languages are typically classified into high-level and low-level programming languages \cite{cormen2022introduction}. The choice of language significantly impacts the performance of the parallel program; low-level languages are usually preferred for parallel programs as they are closer to the hardware with limited abstractions \cite{dean2008mapreduce}. That does not mean that high-level languages cannot write parallel programs, but if speed and efficiency are paramount, it is more appropriate to write a parallel program in a low-level language, just as if you want an efficient serial program, it is crucial to write the program in a language closer to hardware, the same principles apply to parallel programs \cite{herlihy2020art}.

\vspace{0.3cm}
Libraries used for parallel programming also play a crucial role in an application's performance \cite{kulkarni2007optimistic}. Some libraries are meticulously optimized to extract every bit of efficiency from each line of code—employing techniques like low-level synchronization, vectorization, and loop unrolling—to maximize speed on specific hardware \cite{hennessy2011computer}. In contrast, while generally efficient, other libraries may not be as finely tuned. Moreover, even a well-optimized library can exhibit varying performance across standard implementations due to differences in system architectures, compiler optimizations, or algorithmic choices \cite{quinn2004parallel}. Therefore, selecting the appropriate library involves balancing factors such as raw performance, compatibility with target hardware, and overall ease of integration with the existing codebase \cite{fu1997space}.

\vspace{0.3cm}

We implemented Conway's Game of Life as a parallel cellular automaton using MPI in Python and C to compare different programming languages' performance and parallelization capabilities. We ran these implementations on the Cirrus supercomputer at the University of Edinburgh to analyze how each language affects program performance. The program was parallelized using 4 processes.

\begin{figure}[h]
    \centering
    \includegraphics[width=0.5\textwidth]{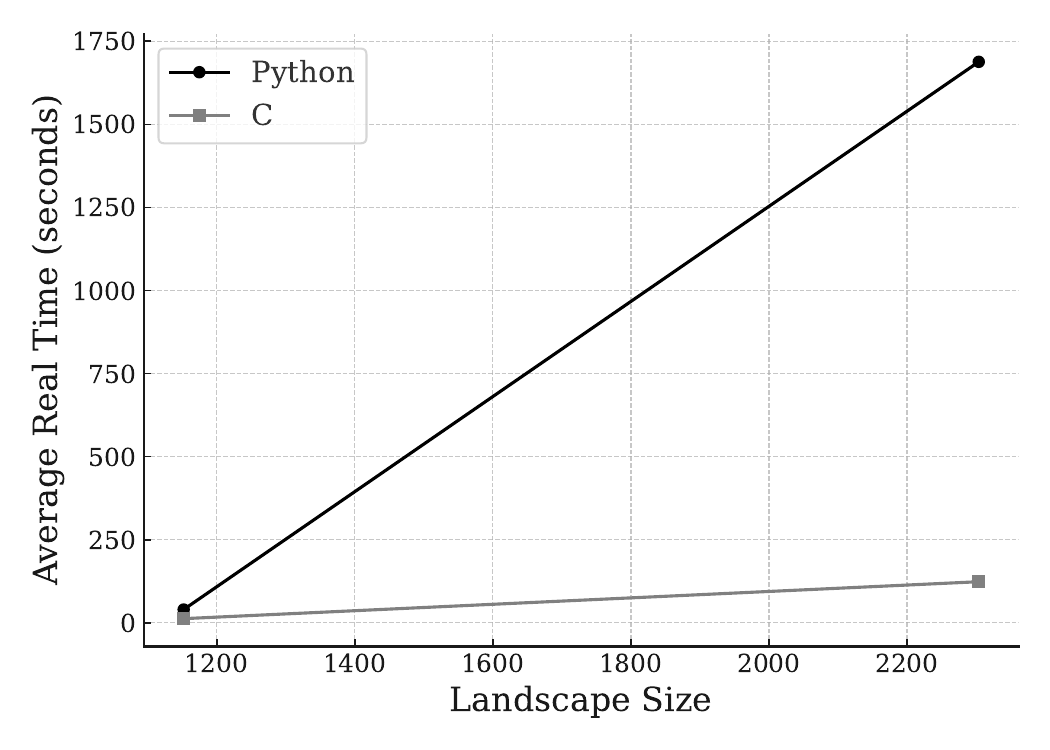}
    \caption{Timing Results of a Cellular Automaton parallelized using Message Passing Interface}
    \label{fig:example}
\end{figure}

Based on the figure, it is evident that C is far more efficient than Python in writing MPI programs. Our experiments were conducted on two landscape sizes, namely 1152 and 2304, and the results clearly indicate that the efficiency gap between the two languages grows as the problem size increases.

\vspace{0.3cm}

One fundamental reason for this disparity is the intrinsic difference in how the two languages are executed. C is a compiled language, which means that its code is translated directly into machine language before execution. This allows for low-level optimizations, efficient memory management, and direct access to hardware resources. Such features are crucial in high-performance computing environments, where every millisecond counts \cite{javier2023performance}. In contrast, Python is an interpreted language that introduces an additional layer of abstraction. This results in runtime overhead due to dynamic type checking, garbage collection, and the interpretation process itself. As the problem size scales up, these overheads become increasingly significant, thereby exacerbating the performance gap \cite{park2020mpipython}.

\vspace{0.3cm}

Furthermore, when it comes to MPI programming, the efficiency of communication between processes is paramount. C-based MPI implementations are typically optimized to leverage the full capabilities of the underlying hardware, including low-latency networks and high-throughput interconnects. Python, although supported by libraries such as \texttt{mpi4py}, adds an extra layer of abstraction over the native MPI calls. This extra layer can introduce latency and additional computational cost, which in turn diminishes its performance, especially in larger-scale problems where communication costs dominate \cite{elshazly2020performance}.

\vspace{0.3cm}

Additionally, as the landscape size doubles, the amount of data to be processed and communicated increases significantly. This not only increases the computational load but also magnifies any inefficiencies in the programming model. In C, the low-level control and static typing allow the program to scale more gracefully under increased load. Python, on the other hand, may suffer from scaling issues due to its inherent overhead, making it less suitable for very large problem sizes in MPI contexts \cite{dalcin2011parallel}.

\vspace{0.3cm}

The combination of compile-time optimizations, efficient memory management, and direct hardware interfacing gives C a substantial advantage over Python for parallel programming. These advantages become even more pronounced as the problem size increases, leading to a widening efficiency gap between the two languages \cite{zosimov2023optimizing}.

\subsection{Hardware}
This section discusses the impact that hardware has on parallel computing. Hardware is broad and overarching, so this section is divided into different components, including the cache, memory, CPU, accelerator, architecture, and interconnects. Each of these elements plays a critical role in ensuring efficient parallel computation. The cache minimizes data access latency by storing frequently used data close to the processor, thereby improving execution speed \cite{tinetti2013cache}. The memory subsystem provides the bandwidth and capacity to manage large volumes of data efficiently, which is crucial for high-performance computing (HPC) applications \cite{fox1991parallel}. The CPU is the central orchestrator, coordinating multiple threads and processes to maximize computational throughput \cite{cheptsov2013parallel}.

\vspace{0.3cm}

Accelerators, such as GPUs and FPGAs, enhance performance by offloading specialized tasks and enabling massive parallelism. Adopting accelerators in HPC systems has become increasingly common due to their ability to execute data-parallel workloads efficiently \cite{bientinesi2015accelerator}. In the subsequent sections, we delve deeper into these components, exploring their architectures, interactions, and how they collectively contribute to the performance and scalability of parallel computing systems.

\subsubsection{CPU}
The CPU (Central Processing Unit) is the primary computing engine responsible for executing instructions and managing data flow. Modern CPUs often feature multiple cores, enabling the simultaneous execution of multiple threads and enhancing computational efficiency \cite{khan2012multi}. Advanced techniques like pipelining, branch prediction, and out-of-order execution further optimize instruction throughput by minimizing stalls and ensuring efficient resource utilization \cite{dezhgosha1996cpu}.

\vspace{0.3cm}

In homogeneous architectures, where all CPU cores are identical, parallel workloads benefit from predictable performance scaling and simplified task scheduling. Uniform core performance ensures efficient load balancing, making such designs ideal for applications like scientific simulations, financial modeling, and numerical computing, where tasks can be evenly distributed across cores \cite{amoretti2020homogeneous}. However, homogeneous CPUs struggle with diverse workloads, lacking the architectural flexibility to optimize specialized tasks such as deep learning inference or high-speed cryptographic processing.

\vspace{0.3cm}

Conversely, heterogeneous CPU architectures combine general-purpose cores with specialized cores or accelerators. This hybrid approach is increasingly popular in modern computing systems, mobile processors, and AI-driven architectures, where high-performance cores manage control flow while energy-efficient cores or domain-specific units handle specialized tasks \cite{bientinesi2015accelerator}.

\vspace{0.3cm}

For efficient parallel computing, CPU architectures must support high-speed inter-core communication, low-latency memory access, and dynamic task scheduling. In multi-core CPUs, cache coherence mechanisms (e.g., the MESI protocol) ensure consistency across cores but can introduce performance overhead due to increased synchronization traffic. Optimizations like NUMA-aware memory placement and thread affinity help reduce cache contention and improve locality, directly enhancing parallel performance \cite{fox1991parallel}.

\subsubsection{Cache}

Caches are small, high-speed memory units situated close to the CPU cores. They store frequently accessed data and instructions to reduce latency and improve performance. Typically, modern processors employ a multilevel cache hierarchy (L1, L2, and sometimes L3), each balancing speed and capacity \cite{fox1991parallel}.

\vspace{0.3cm}

In parallel computing, efficient cache utilization is critical as multiple cores require fast access to shared data. Poor cache management can lead to contention and increased memory access latencies, reducing system efficiency \cite{tinetti2013cache}.

\vspace{0.3cm}

A key challenge in multi-core and many-core systems is cache coherence, which ensures that all processor cores see a consistent view of memory. Cache coherence protocols, such as MESI (Modified, Exclusive, Shared, Invalid), are crucial in synchronizing shared data across multiple caches. However, maintaining coherence introduces performance overhead, as frequent invalidations and updates can increase memory traffic, affecting parallel efficiency \cite{khan2012multi}. To mitigate this, modern architectures incorporate non-uniform cache architectures (NUCA) and directory-based coherence mechanisms to balance latency and bandwidth.

\subsubsection{Memory}

Memory, commonly implemented as DRAM, is the primary workspace for data and instructions during program execution. The performance of the memory subsystem—in terms of capacity, bandwidth, and latency—directly influences the efficiency of parallel applications. For instance, distributed memory architectures require efficient data communication strategies to optimize performance \cite{kimpe2006memory}.

\vspace{0.3cm}

In heterogeneous systems, managing distinct memory spaces for different types of processing units introduces complexities in data coherence and efficient access. Consequently, optimizing memory access patterns is key to improving performance \cite{jong2010memory}. Coordination and synchronization are crucial for optimal performance, as the limited uniformity of the architecture might cause unpredictability in the performance of parallel programs.

\vspace{0.3cm}

One of the major bottlenecks in parallel computing is memory bandwidth limitations. As the number of processing cores increases, the demand for memory access grows, potentially leading to memory contention and bottlenecks in shared-memory architectures. High-bandwidth memory (HBM) and DDR5 are designed to mitigate these issues by offering increased bandwidth and reduced latency \cite{bientinesi2015accelerator}.

\subsubsection{Accelerator}

Accelerators have become prevalent in parallel computing due to their ability to handle specialized tasks that demand high computational throughput and parallel data processing. Unlike traditional CPUs, which prioritize general-purpose computing and sequential task execution, accelerators exploit task and data parallelism by leveraging architectures optimized for specific workloads. For instance, GPUs (Graphics Processing Units) utilize a Single Instruction Multiple Thread (SIMT) execution model, making them particularly effective for matrix operations, convolutional computations in deep learning, and large-scale simulations in computational physics and molecular dynamics \cite{bientinesi2015accelerator}.

\vspace{0.3cm}

Similarly, FPGAs (Field-Programmable Gate Arrays) offer customizable hardware acceleration, allowing direct circuit-level optimization of specific tasks. Unlike fixed-function GPUs, FPGAs provide fine-grained parallelism and low-latency execution, making them suitable for real-time data processing, signal processing, and cryptographic applications \cite{jong2010memory}.

\vspace{0.3cm}

On the other hand, ASICs (Application-Specific Integrated Circuits) provide the highest efficiency level for targeted applications, such as AI inference, blockchain computations, and high-frequency trading. By being hardwired for specific functions, ASICs eliminate overhead associated with general-purpose processing, leading to unmatched power efficiency and computational density \cite{hao2023ai}.

\vspace{0.3cm}

While accelerators significantly enhance parallel computing capabilities, their integration into heterogeneous architectures introduces multiple challenges. Efficient memory management becomes critical, as different processing units operate on separate memory spaces. Unified Memory Architecture (UMA), zero-copy memory transfers, and direct memory access (DMA) aim to bridge these gaps, ensuring minimal data transfer overhead \cite{fox1991parallel}.

\subsubsection{Architecture}

The architecture of a computing system defines the overall design and organization of its hardware components, including the CPU, cache, memory, accelerators, and interconnects. The choice of architecture directly influences parallel computing performance by determining computational efficiency, scalability, energy consumption, and ease of programming. A key distinction in modern architectures is between homogeneous and heterogeneous designs, which have unique implications for parallel computing performance \cite{amoretti2020homogeneous}.

\vspace{0.3cm}

Homogeneous architectures provide predictable performance scaling and efficient task scheduling, while heterogeneous architectures combine different processing units, such as general-purpose CPUs with specialized GPUs, FPGAs, or TPUs, to leverage their strengths \cite{bientinesi2015accelerator}.

\subsubsection{Interconnects}

Interconnects are the communication backbone linking various hardware components within a computing system, such as CPUs, memory, and accelerators. Their performance is crucial in parallel computing, as they directly impact data transfer speed, latency, and overall system efficiency \cite{dezhgosha1996cpu}.

\vspace{0.3cm}

One of the primary challenges in parallel computing is minimizing communication overhead while maximizing data locality and concurrency. High-speed, low-latency interconnects—such as PCIe, NVLink, and Intel's Compute Express Link (CXL)—are designed to facilitate high-throughput data transfer between CPUs, GPUs, and accelerators \cite{khan2012multi}.

\section{Tooling and Ecosystem}

Numerous tools aid parallel programming, and the number has been growing since its inception. These tools are enveloped within a rich ecosystem that helps write, debug, and optimize concurrent applications \cite{mattson2004patterns, topcuoglu2002performance}. In this section, we explore the various components of this ecosystem, from foundational programming frameworks to performance analysis tools, simulation platforms, and community resources that together empower developers to build robust parallel systems.

\subsection{Programming Frameworks and Libraries}

A wide array of programming frameworks and libraries forms the spine of parallel development. Widely adopted frameworks such as MPI (Message Passing Interface) and OpenMP provide standardized approaches to parallelism on distributed-memory and shared-memory systems, which have been discussed earlier. There are specialized libraries like CUDA and OpenCL, which haven't been mentioned, that target the massive parallelism offered by GPUs, enabling developers to harness the power of heterogeneous architectures for compute-intensive tasks fully \cite{gropp1999using, dagum1998openmp}.

\vspace{0.3cm}

These frameworks offer the fundamental constructs for process synchronization, task distribution, and communication and integrate with modern programming languages like C++, Python, and Rust to provide a high degree of abstraction. These abstractions enable developers to focus on algorithmic design rather than low-level details while achieving efficient execution across multi-core and many-core environments.

\subsection{Performance Analysis and Debugging Tools}

Identifying bottlenecks and ensuring the correctness of parallel applications requires a robust suite of profiling and debugging tools. Tools such as Intel VTune and NVIDIA Nsight offer deep insights into applications' runtime behavior, including CPU and GPU utilization, memory bandwidth usage, and communication overheads \cite{intel2022vtune, nvidia2022nsight}. These tools are essential for fine-tuning performance, as they help pinpoint areas where parallel efficiency may be improved.

\vspace{0.3cm}

In parallel environments, debugging challenges such as race conditions, deadlocks, and synchronization issues become more pronounced, including root cause analysis; when parallelizing a serial program, it moves the number of instances your program is running from singular to multiple, which makes it more complicated in trying to identify where the issue is when a bug arises. Specialized debugging tools like TotalView and Allinea DDT are designed to handle the complexity of multi-threaded and multi-process applications. They allow developers to step through concurrent executions, inspect thread states, and monitor inter-process communication, ensuring that parallel programs run efficiently and correctly \cite{rogers2003totalview}.

\subsection{Simulation and Visualization Platforms}

Simulation and visualization tools play a crucial role in the design and analysis of parallel systems. These platforms allow developers to model and predict the behavior of complex systems under various workloads, providing a sandbox environment for testing theoretical models like PRAM, BSP, and LogP \cite{valiant1990bridging, culler1996logp}. This helps them understand the dynamics of their parallel programs before even writing code. By simulating parallel architectures and workloads, these tools provide a robust environment for analyzing the scalability and synchronization overhead before deployment on actual hardware.

\vspace{0.3cm}

Visualization also plays a critical role; graphing tools allow the process to assist in interpreting performance data, which can be critical in benchmarking and understanding implementation dynamics to make better decisions about parallel programs, and it also makes it easier to understand complex interactions within parallel systems. Graphical representations of thread activity, memory usage, and interconnect traffic can highlight inefficiencies that will not be visible simply by compiling the program and executing the code. Visualization has become very critical and valuable to the parallel computing ecosystem because of the insights that lead to more efficient and resilient parallel systems \cite{perumalla2006parallel}.

\subsection{Community, Documentation, and Learning Resources}

The active communities surrounding parallel programming frameworks and tools greatly enhance the tooling ecosystem. Open-source projects, online forums, and collaborative platforms such as GitHub provide an environment where developers can share code, report issues, and contribute to improvements \cite{raymond1999cathedral}. This collaborative spirit accelerates innovation and ensures that best practices are disseminated widely across the community \cite{von2003democratizing}.

\vspace{0.3cm}

In addition to community support, comprehensive documentation, tutorials, and training courses are critical for empowering developers to use these tools effectively. Many frameworks offer extensive official documentation \cite{openmp2021spec}, while academic institutions and industry leaders provide webinars, workshops, and online courses to help new and experienced developers \cite{smith2011openmp}. The continuous evolution of these resources ensures that computer scientists remain up-to-date with the latest advancements and techniques in parallel programming, fostering an environment of lifelong learning and innovation \cite{blelloch1996programming}.

\subsection{Integration and Deployment Tools}

Integration with modern build systems and deployment tools is essential to bringing parallel applications from development to production. Tools such as CMake, Make, and various IDE plugins facilitate the compilation and linking of parallel codebases across different platforms \cite{hoffman2016cmake}. Furthermore, containerization technologies like Docker and orchestration tools like Kubernetes are increasingly used to deploy parallel applications in scalable, cloud-based environments, ensuring consistent performance and manageability \cite{bernstein2014containers}.

\vspace{0.3cm}

Deployment tools also play a critical role in monitoring and maintaining the health of parallel applications once they are in production. Real-time monitoring systems, log aggregators, and automated testing frameworks help ensure that performance remains optimal and that any issues are promptly identified and addressed \cite{liu2017automated}. This holistic approach to tooling—from development to deployment—ensures that parallel applications are robust, scalable, and ready to meet the demands of modern computing workloads \cite{hennessy2011computer}.

\section{Case Study of a Road Traffic Simulation}

Using a road traffic simulation as a case study, we identify both suitable and unsuitable parallel programming patterns for implementation. We then analyze how different hardware configurations and programming languages affect these patterns' performance. Additionally, we present methods to evaluate the effectiveness of the program implemented with the selected pattern. The detailed specifications of the road traffic simulation are provided in the appendix.

\subsection{Chosen Design Patterns}

\subsubsection{Geometric Decomposition}

\paragraph{Overview of Geometric Decomposition}

Geometric or domain decomposition is a parallel computing strategy that partitions a spatial domain into smaller, manageable subdomains \cite{michalakes1999data}. It could be suited for the simulation model but not without drawbacks; it is typically used in domain problems that require localized handling and communication between neighbors, often in conjunction with the Message Passing Interface \cite{smyk2011fdtd}. The road traffic simulation represents roads and junctions as a graph structure, where vehicles move from one node to another, engaging with local features such as traffic lights and possibly navigating events such as crashes and collisions. By leveraging geometric decomposition, the simulation can be partitioned based on the geographical layout of the road network, allowing for efficient parallel processing and management of localized interactions. Still, the partitioning is not as straightforward, and the simulation dynamics complicate utilizing the strategy \cite{garmann1997dynamic}.

\vspace{0.3cm}

\noindent It is essential to state that the data structure we are dealing with for this simulation is a graph. This makes it challenging to decompose its subdomains across multiple processing elements without problems in synchronization and efficient partitioning. Unlike linear data structures such as arrays or linked lists, which can be easily divided into chunks and distributed based on the amount of UEs available, graph structures involve complex interconnections that complicate decomposition \cite{bourgeois2017gis}. At the same time, tools like MPI Graph Topology can assist in decomposing a graph for parallel computation; there are bottlenecks in using this approach specifically for this simulation model \cite{paulius2015scalability}.

\vspace{0.3cm}

\noindent However, dividing the number of elements by the number of processes and distributing the remainder across processing elements is not straightforward for a graph data structure. A graph operates on vertices and edges, not on a simple linear sequence of elements, making it difficult to partition evenly \cite{moulitsas2005graph}. Although the unidirectional graph-like model for the road simulation can simplify decomposition somewhat, the underlying complexities of vertex and edge distribution remain significant. Graphs consist of vertices connected by edges, and these connections often span across different partitions when the graph is divided. This results in communication and synchronization overhead as vertices in one partition may frequently interact with vertices in another. Minimizing such inter-partition edges is crucial for reducing communication costs and enhancing performance, but achieving this balance is inherently challenging due to the graph's topology \cite{pellegrini2011current}.

\vspace{0.3cm}

\noindent Despite these challenges, it is possible to decompose a graph and distribute it across processing elements. Graph partitioning has become a well-established technique in parallel computing, with various strategies developed to address its complexities \cite{engelen2000graph}. Methods such as adjacency lists are commonly used to enable efficient graph processing, particularly for sparse graphs like road networks. Additionally, graph partitioning libraries provide multiple techniques for dividing graphs among processing elements (PEs). One of the most widely used tools for this purpose is the sequential partitioner METIS \cite{karypis2000graph}. Parallel partitioners like ParMETIS, PT-Scotch, and JOSTLE are also designed explicitly for distributed-memory architectures in parallel programs. However, these implementations have specific challenges, which we will explore in detail \cite{predari2016graph}.
\vspace{0.3cm}

\paragraph{Advantages in Applying Geometric Decomposition for this Road Traffic Simulation Model}

As discussed earlier, the road network's graph can be naturally divided into subgraphs or regions, each containing a subset of junctions and connecting roads. Since vehicles interact primarily with their immediate environments, such as the road they are on or the intersection they are approaching—most of the computational workload is localized within these subdomains and can be effectively managed \cite{miller2021theoretical}. Tools like neighborhood collective operations can be used to manage communications in the topology if implemented effectively using Message Passing Interface. For instance, calculating a vehicle's position, speed adjustments due to road congestion, and decisions at junctions are all confined to specific areas of the network; this is possible, but the simulation dynamics could typically complicate this. However, we have to discuss this as the localized processing of sub-domains is one of the highlights of domain decomposition due to its ability to be very efficient as new processes are added. Geometric decomposition could capitalize on this locality by assigning each subdomain to a different processor, enabling concurrent communication and updates without significant interference \cite{svitenkov2016partitioning}.

\vspace{0.3cm}

Geometric decomposition can enhance computational efficiency and scalability in huge problem sizes but depends on the distribution's granularity \cite{kirmani2018scalability}. Parallel processing of subdomains means that the simulation can handle more vehicles and more extensive road networks without a linear increase in computation time; this would allow the program to scale proportionally to the number of available PEs, but there would also be limitations as processes grow, due to the serial portion of the code that becomes a bottleneck in parallelization. Each processor could focus on its localized setting, handling interactions in that instance. This division of labor reduces the computational overhead. It allows for more frequent updates and finer-grained simulation details within each subdomain. Still, it is essential to note that allowing this is not as straightforward as discussed due to dynamism and the concurrent activities within the simulation, making it difficult for each execution unit to manage its domain effectively while also being part of the global environment.

\vspace{0.3cm}

Moreover, geometric decomposition facilitates effective management of the simulation's data structures and memory usage. By keeping data localized to specific subdomains, the simulation minimizes the need for global data synchronization and reduces memory contention among processors \cite{axner2008performance}. This localization is particularly beneficial for recording and reporting simulation statistics when needed as it would allow calculations and derivations of nuances within the simulation, which might be challenging to capture in other parallel pattern environments such as the Actor model due to the granularity as it would require that each actor sends a message as an update and that can impact performance due to message passing overload. Each processor can independently maintain and update these statistics for its subdomain in the Geometric pattern, contributing to a comprehensive final report without excessive inter-processor communication.
\vspace{0.3cm}

\paragraph{Challenges in Applying Geometric Decomposition for this Road Traffic Simulation Model}

However, applying geometric decomposition requires careful consideration of inter-subdomain interactions and load balancing \cite{guo2016application}. Vehicles moving between subdomains introduce the need for communication between processors to ensure seamless transitions and consistent simulation states. There must be established to handle these boundary conditions without introducing significant latency; vehicles transferring locations between processed would have to be communicated, and as the number of cars increases and as the simulation scales with additional complexity, this can eventually become a challenge and a bottleneck in the simulation. The road network may have varying complexity and traffic density regions, leading to unequal computational loads across processors. Dynamic load balancing would required, which is not only complex to implement but also complex to manage \cite{goodrich2016parallel}.

\vspace{0.3cm}

Another significant challenge lies in the limitations of graph partitioning libraries. The graph partitioning libraries are typically optimized for large-scale graphs with millions of vertices and edges \cite{chamberlain2001graph}. The specifications for the road network provided would not typically be optimized for graph partitioning libraries like METIS, which usually add overhead. Graph partitioning libraries use sophisticated algorithms to segment the graph into efficient, adequate proportions. The time taken to partition a graph increases with the size and complexity of the graph. For massive graphs, graph partitioning libraries are optimized to handle them efficiently. Still, for smaller graphs, the relative partitioning time might be significant compared to the overall simulation runtime, which would be evident in the road traffic simulation, eventually making the program inefficient \cite{zheng2023parallel}. If partitioning your graph takes hours and your simulation runs for minutes, it is not the most efficient scenario; a better approach would be a better ratio in proportion to the simulation time.

\vspace{0.3cm}
\noindent Furthermore, graph partitioning libraries, such as METIS's algorithms, have specific time complexities \cite{karypis1998fast}. The multilevel approach typically operates in near-linear time relative to the number of edges, but the constants involved can be non-trivial. Additional memory is required to store intermediate graph representations during the coarsening and partitioning phases. This includes data structures for the coarsened graphs, mappings between levels, and other auxiliary information \cite{karypis1999multilevel}.

\vspace{0.3cm}

\noindent While for massive graphs, this memory overhead is manageable and often justifiable by the performance gains in partitioning, the relative memory consumption might be less efficient for smaller graphs, potentially limiting resources for other simulation components \cite{chen2007partitioning}.

\vspace{0.3cm}

It's important to note that the simulation dynamics do not favor the linear pattern of geometric decomposition. For example, the problem's specifications stated that vehicles followed recalculated routes; if this weren't the case, this design strategy would have been more viable as we could apply data parallelism to represent each road as a queue and distribute partitions across processes. The overhead would be more tolerable, but a critical component for this road traffic model is when vehicles arrive at a junction, their destination can be re-routed, which would make readjusting the queue impractical and further exacerbate the unsuitability of this approach for the road traffic simulation \cite{bader2006parallel}.

\vspace{0.3cm}

Crashes at junctions without traffic lights introduce another layer of complexity in a partitioned environment. Since crashes are localized events involving vehicles from different partitions, detecting and managing these incidents requires robust mechanisms to prevent race conditions \cite{hoefler2010performance}. What happens when two cars crash in two distinct processes simultaneously? How do you account for it in the simulation? It would typically require managing a synchronization procedure, which would likely harm performance due to its need to be aware at every point in the simulation. UEs must collaborate to update the state of junctions involved in crashes, ensuring that the removal of vehicles from the simulation is handled accurately and consistently across all relevant partitions \cite{rauber2013parallel}. This necessity for coordinated event handling can increase the computational and communication overhead, challenging the overall efficiency of the simulation.

\vspace{0.3cm}

Fuel consumption and vehicle removal processes also experience significant impacts from graph partitioning. Each UE is responsible for tracking the fuel levels of vehicles within its partition, necessitating communication with other UEs when vehicles run out of fuel or are removed due to crashes. Efficiently managing these dynamic state transitions across partitions with the appropriate data structures and memory buffers would require careful programming \cite{grama2003introduction}. The inter-PE communication necessary for these operations can introduce additional latency, particularly when vehicle states frequently change or interact across multiple partitions \cite{tanenbaum2007distributed}.

\vspace{0.3cm}

It is also possible that domain decomposition could alter vertex ordering within the graph. Since graph partitioning libraries usually focus on optimizing load balancing and reducing communication overhead, they rearrange the vertices to achieve these goals \cite{karypis1998fast}. In the context of road traffic simulation, where junctions represent critical points for vehicle movements and traffic light operations, this reordering can complicate the management of vehicle routes and traffic flow. Vehicles moving between partitions require seamless communication between UEs to accurately update their locations and statuses \cite{hennessy2017computer}. This necessity introduces latency and demands robust synchronization mechanisms to ensure vehicle movements and traffic light states remain consistent across all partitions.

\vspace{0.3cm}

The geometric decomposition pattern complicates the dynamic addition of vehicles, primarily due to challenges like identifying the appropriate unit of execution (UE) for placing new vehicles, tracking newly added cars, and managing the communication overhead involved in dynamically handling the addition and removal of vehicles \cite{quinn2004parallel}. As a result, this strategy is not well-suited for supporting this specific dynamic aspect of the simulation.

\vspace{0.3cm}

These points all emphasize one theme within the simulation that geometric decomposition cannot possibly handle: synchronization to manage the simulation's dynamism and nonlinearity \cite{hoefler2010performance}. In the next section, we will discuss the synchronization required.

\vspace{0.3cm}

\begin{figure}
\centering
\begin{tikzpicture}[scale=0.5]
    % Define colors
        \colorlet{p1color}{blue!15}
        \colorlet{p2color}{green!15}
        \colorlet{p3color}{orange!15}
        \colorlet{junctioncolor}{black!45}
        \colorlet{roadcolor}{gray!30}
        \colorlet{carcolor}{yellow!70!black}
        \colorlet{boundarycolor}{purple!70}
        
    % Draw partitions
    \fill[p1color] (0,0) rectangle (4,8);
    \fill[p2color] (5,0) rectangle (9,8);
    \fill[p3color] (10,0) rectangle (14,8);
    \draw[dashed, gray!60] (0,0) rectangle (4,8);
    \draw[dashed, gray!60] (5,0) rectangle (9,8);
    \draw[dashed, gray!60] (10,0) rectangle (14,8);
    
    % Draw roads
    \fill[roadcolor] (0.5,2) rectangle (3.5,2.5);
    \fill[roadcolor] (0.5,5.5) rectangle (3.5,6);
    \fill[roadcolor] (5.5,2) rectangle (8.5,2.5);
    \fill[roadcolor] (5.5,5.5) rectangle (8.5,6);
    \fill[roadcolor] (10.5,2) rectangle (13.5,2.5);
    \fill[roadcolor] (10.5,5.5) rectangle (13.5,6);
    
    % Draw junctions
    \foreach \x/\y in {0.5/1.5, 3.5/1.5, 0.5/6.5, 3.5/6.5, 
                       5.5/1.5, 8.5/1.5, 5.5/6.5, 8.5/6.5,
                       10.5/1.5, 13.5/1.5, 10.5/6.5, 13.5/6.5} {
        \fill[junctioncolor] (\x,\y) circle (0.3);
    }
    
    % Draw cars
    \fill[carcolor] (1,2.15) rectangle (1.5,2.35);
    \fill[carcolor] (6,5.65) rectangle (6.5,5.85);
    \fill[carcolor] (11,2.15) rectangle (11.5,2.35);
    
    % Draw partition boundaries
    \draw[boundarycolor, thick] (4.5,0) -- (4.5,8);
    \draw[boundarycolor, thick] (9.5,0) -- (9.5,8);
    
    % Labels
    \node[align=center] at (2,-0.5) {Partition 1};
    \node[align=center] at (7,-0.5) {Partition 2};
    \node[align=center] at (12,-0.5) {Partition 3};
    
    % Legend
    \node[anchor=west] at (15,7) {\textbf{Legend:}};
    \fill[junctioncolor] (15,6) circle (0.2) node[right=0.3cm] {Junction};
    \fill[roadcolor] (15,5) rectangle (15.4,5.2) node[right=0.5cm] {Road};
    \fill[carcolor] (15,4) rectangle (15.4,4.2) node[right=0.5cm] {Vehicle};
    \draw[boundarycolor, thick] (15,3) -- (15.4,3) node[right=0.5cm] {Boundary};

\end{tikzpicture}
\caption{Domain Decomposition in Road Traffic Simulation}
\label{fig:domain_decomposition}
\end{figure}
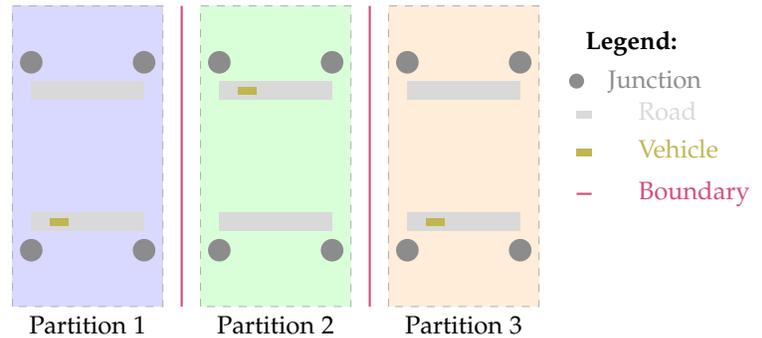

\vspace{0.3cm}

\paragraph{Synchronization and Memory Considerations}

The synchronization mechanisms needed for this parallel design strategy in the road traffic simulation are complex and robust, making their development, testing, and debugging time-consuming \cite{rauber2013parallel, grama2003introduction}. A significant bottleneck is ensuring that each partition remains independent while the overall model operates seamlessly.

\vspace{0.3cm}

Memory requirements are generally lower per processing element in distributed memory architectures than in shared memory architectures \cite{tanenbaum2007distributed}. However, overall memory usage remains a consideration because the data structures used to store graph representations still consume significant memory resources. The most considerable bottleneck associated with this design pattern is latency, arising from the communication overhead between distributed processing elements. Multiple messages are being sent between PEs simultaneously to ensure synchronization, which can lead to a build-up in the queues, further exacerbating latency and affecting the overall time of the simulation \cite{kumaran2003performance}.

\vspace{0.3cm}

Due to the reordering of vertices, messaging patterns across processing elements (PEs) can become unpredictable. This unpredictability hampers traceability, root cause analysis, and debugging processes, limiting the ability to successfully and efficiently derive an optimal and correct simulation within an adequate amount of time \cite{hoefler2010performance}. When vertices are distributed in a non-sequential manner, tracking the flow of messages and identifying where and why specific issues occur becomes challenging. This lack of clarity complicates the debugging process, making it difficult to pinpoint errors or inefficiencies within the simulation, eventually leading to significantly larger development time.

\vspace{0.3cm}

Beyond the challenges in messaging and debugging, vertex disordering adversely affects data locality and cache performance \cite{hennessy2017computer}. In a well-ordered graph, related vertices are often stored closely in memory, facilitating faster data access and improved cache utilization. However, when vertices are reordered and dispersed across different partitions, the simulation may experience increased cache misses and slower data retrieval times. This degradation in cache performance can significantly reduce the overall efficiency of the simulation, as more time is spent accessing scattered data rather than processing it; this can also be very time-consuming as trying to find the next edge in the graph non-linearly might not be very efficient. Additionally, the complexity of data access patterns increases, making it harder to optimize memory usage and further impacting the simulation's performance \cite{topcuoglu2002performance}.

\vspace{0.3cm}

\paragraph{Impact on Hardware}

Multi-core or many-core parallelism is typically integrated into general-purpose CPUs. In contrast, Graphical Processing Units (GPUs) offer a high degree of parallelism through numerous simple cores. However, effectively using GPUs requires structuring compute problems to fit GPU hardware's regular, parallel nature \cite{kirk2016programming}. Additionally, the simulation in question would involve irregular operations that are “sparse”—entailing numerous random memory accesses due to the nonlinearity of the simulation—which can negatively impact data locality and cache performance \cite{cuda2019programming}.

\vspace{0.3cm}

While shared-memory systems are adequate for the scale of this simulation, multi-node systems are more suitable for handling the I/O required at the simulation's end. Multi-node systems, especially those with NUMA (Non-Uniform Memory Access) architecture, benefit from data decomposition, which allows more efficient data placement across memory banks \cite{kimpe2006memory}. If the simulation is properly load-balanced and inter-unit communication is minimized, these systems can better use the available memory bandwidth.

\vspace{0.3cm}

\paragraph{Impact on Programming Language}
\noindent For implementing this program, I recommend using a low-level language such as C, C++, or Fortran, with a particular preference for C. C offers fine-grained control over memory management and is well-supported by libraries for geometric decomposition, making it ideal for this task \cite{karniadakis2003parallel}. While high-level languages like Python are an option, they are less practical for efficiency due to the overhead involved with recursive calls and garbage collection. Python acts as an abstraction over C, meaning that using it introduces extra overhead, lacks compilation advantages, and makes it more challenging to optimize the program, ultimately exacerbating execution speed and inefficiency \cite{misra1994powerlist}.

\vspace{0.3cm}

\paragraph{Summary and Transition to Alternative Strategies}
\noindent In summary, geometric decomposition is a well-tested and foundational design pattern in parallel computing; it allows flexibility and imposes no limitation on how granular the problem can be decomposed, nor on how many PEs the load can be distributed across \cite{quinn2004parallel}. While it is ideal for linear data structures like arrays, linked lists, and structures that follow a sequential order, it could also be suitable for a graph data structure. However, the specifications and dynamics of this problem limit its potential. Consequently, this leads us to explore another parallel design strategy that would be ideal for this problem—the Actor Model—which will be discussed in the next section \cite{hewitt1973actor}.

\subsubsection{Actor Model}

\paragraph{Overview of the Actor Model}
Given the unpredictability and nonlinearity of computations in road traffic simulations, we propose using the actor model rather than employing a decomposition strategy to solve the problem. This approach is a mathematical theory that defines 'Actors' as the fundamental building blocks of concurrent computation \cite{hewitt2015actor}. True to its name, the model involves independent entities (actors) that leverage the nonlinearity of the simulation. While it doesn't follow the traditional parallel computing design patterns, the Actor model has gained popularity for its ability to handle high levels of concurrency. In this framework, an actor is a computational entity that responds to messages it receives \cite{hewitt2015actor}.

\vspace{0.3cm}

\paragraph{Rationale for Choosing the Actor Model}
This model introduces flexibility and dynamism that wasn't achievable with geometric decomposition. The unpredictability of partition placements across execution units (UEs) is no longer an issue. The Actor model allows actors to be mapped to PEs in the programmer's chosen order. Still, significant considerations should be factored in when mapping to optimize for efficiency. However, it's important to note that the graph structure becomes irrelevant and redundant within this framework. All components within the simulation can act independently as actors. This includes the roads, junctions, and vehicles; these actors would be concurrent entities communicating through asynchronous message-passing, which imposes no limitation on the scalability of this program and allows the company to run even heavier simulations due to the amount of concurrency that can be extracted from modeling the simulation using the Actor model \cite{haller2012actors}.

\vspace{0.3cm}

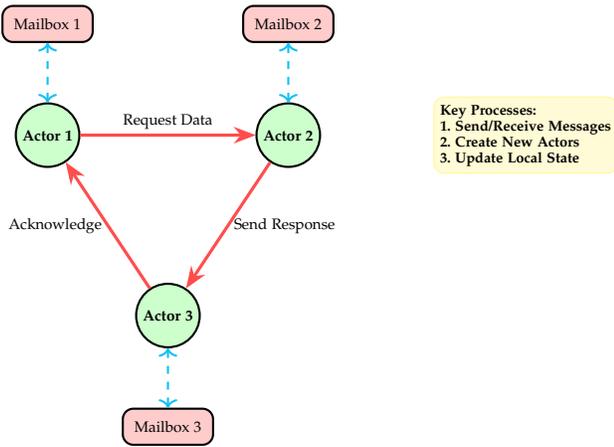
\begin{figure}[h!]
\centering
\begin{tikzpicture}[
    scale=0.8,  % Reduced scale for smaller figure
    every node/.style={scale=0.6, text=black},  % Set default text color to black
    actor/.style={
        draw,
        circle,
        minimum size=1.2cm,  % Reduced size
        thick,
        fill=green!20,       % Updated fill color from second figure's 'vehicle'
        text=black,          % Changed from white to black
        font=\bfseries
    },
    message/.style={
        -{Stealth[scale=1.0]},  % Adjusted arrow size
        thick,
        draw=red!70,             % Updated draw color to match second figure's 'message'
        line width=1.2pt
    },
    box/.style={
        draw,
        rectangle,
        minimum width=2cm,        % Reduced width
        minimum height=0.8cm,     % Reduced height
        thick,
        fill=red!20,              % Updated fill color from second figure's 'junction'
        text=black,               % Changed from white to black
        rounded corners
    },
    label/.style={
        font=\small\itshape,
        text width=2.5cm,         % Adjusted text width
        align=left,
        fill=pink!30,             % Kept or updated as per second figure
        rounded corners,
        inner sep=4pt,             % Reduced inner separation
        draw=red!50,              % Updated draw color
        thick,
        text=black                 % Ensured text is black
    },
    mailbox/.style={
        <->,
        thick,
        draw=cyan!70,              % Updated draw color from second figure
        dashed
    },
    annotation/.style={
        font=\small\bfseries,
        fill=yellow!20,            % Kept consistent with second figure's 'summary'
        rounded corners,
        inner sep=4pt,             % Reduced inner separation
        draw=yellow!50,            % Updated draw color
        text=black                 % Changed text color to black
    }
]

% Actors
\node[actor] (A1) at (0,0) {Actor 1};
\node[actor] (A2) at (4,0) {Actor 2};
\node[actor] (A3) at (2, -3) {Actor 3};

% Mailboxes
\node[box, above=0.8cm of A1] (M1) {Mailbox 1};
\node[box, above=0.8cm of A2] (M2) {Mailbox 2};
\node[box, below=0.8cm of A3] (M3) {Mailbox 3};

% Messages
\draw[message] (A1) -- node[above, text=black] {Request Data} (A2);
\draw[message] (A2) -- node[right, text=black] {Send Response} (A3);
\draw[message] (A3) -- node[left, text=black] {Acknowledge} (A1);

% Mailbox Connections
\draw[mailbox] (A1) -- (M1);
\draw[mailbox] (A2) -- (M2);
\draw[mailbox] (A3) -- (M3);

% Annotation Box
\node[annotation, right=1.5cm of A2] (Anno) {
    \begin{tabular}{@{}l@{}}
        \textbf{Key Processes:} \\
        1. \textbf{Send/Receive Messages} \\  
        2. \textbf{Create New Actors} \\     
        3. \textbf{Update Local State}         
    \end{tabular}
};

% Decorative Brace
\end{tikzpicture}
\caption{The Actor Model of Computation}
\label{fig:actor-model-improved}
\end{figure}

\paragraph{Advantages of the Actor Model}

The actor model has significant advantages in terms of concurrency and scalability. This parallelism is crucial for large-scale simulations involving thousands of junctions and vehicles. The model inherently supports high concurrency, allowing numerous actors to execute simultaneously without interfering with each other's states, thus improving overall simulation performance and scalability \cite{hayduk2013speculative}.

\vspace{0.3cm}

Another key benefit is the simplified synchronization and state management. In the Actor model, each actor maintains its state and interacts with others solely through message passing. This isolation minimizes the risks of race conditions and deadlocks. For instance, when a vehicle interacts with a junction, it sends a message requesting permission to proceed, eliminating the need for complex locking mechanisms. This leads to cleaner, more maintainable code and reduces the complexity of managing shared resources \cite{haller2012actors}.

\vspace{0.3cm}

\paragraph{Drawbacks of the Actor Model}

However, there are notable drawbacks to using the Actor model for this simulation. One primary concern is the overhead associated with message passing. The simulation involves frequent interactions between many actors, such as vehicles requesting access to junctions or updating their routes. This high volume of messages can introduce significant latency and processing overhead, potentially slowing down the simulation, especially when dealing with real-time constraints like traffic light synchronization and vehicle movements \cite{scholliers2014parallel}.

\vspace{0.3cm}

The Actor model can also lead to resource contention and bottlenecks, particularly at heavily trafficked junctions. Junction actors may become overwhelmed with incoming messages from numerous vehicles, limiting their ability to process requests efficiently. This can hinder scalability and degrade performance, as the throughput of these critical actors directly impacts the overall simulation. Furthermore, managing many actors can result in high memory consumption, posing challenges for simulations that require hundreds of thousands of actors to model extensive road networks \cite{kim1997thal}.

\vspace{0.3cm}

The Actor pattern can also make traceability difficult, as we might struggle to understand the overall nature of the program. Debugging tools might help, but a measure of concurrency within the program would make it challenging to perform root cause analysis \cite{rosa2016efficient}.

\vspace{0.3cm}

The actor pattern is the preferred pattern over the design strategies due to its ability to handle dynamism; the actor pattern should also be the preferred pattern for a variety of simulation problems; geometric decomposition is a clever and concise parallel design but should be more suited for computations which are predetermined beforehand, although we discussed the limitations of graph partitioning, this is not the most crucial problem of using this strategy the main issues that arise is synchronization and how to write code that cannot be solved in a linear sequence \cite{agha1997parallel}.

\vspace{0.3cm}

\paragraph{Impact on Hardware}

The Actor model profoundly impacts the selection of hardware platforms due to its inherently concurrent and decentralized nature. Hardware that excels in parallel processing and efficient inter-process communication is essential for optimal performance. Multi-core and many-core processors and distributed computing systems are well-suited because they can assign actors to separate processing units, enabling precise concurrent execution \cite{kim1997thal}. Additionally, efficient and high-performance networks would be needed, especially as the scale grows; for a large simulation, writing the program on a consumer-grade system would be very inefficient due to the limited number of cores and also the synchronization of those, as throughput can be a significant determinant of the performance of actor-based systems.

\vspace{0.3cm}

Implementing the Actor model requires a message-passing library, which makes hardware choice a critical factor. Distributed computing architectures like supercomputers are ideal for achieving maximum scalability. While there is no definitive processor for implementing this model—since most modern multi-core processors meet the basic requirements—careful consideration should be given to selecting the most suitable compiler and compiler flags for the architecture to maximize efficiency and also maximize the utility of resources \cite{hiesgen2017opencl}.

\vspace{0.3cm}

\paragraph{Implementation Considerations}

As all actors act independently, we can identify them by their ID; we can map an actor to a process or share the actors across processes and have a hashing procedure to help us understand which process to direct a message to. A hashing procedure maps an actor's unique ID to a specific shard or process by applying a hash function, ensuring that all messages intended for a particular actor are consistently directed to the same process. This method enables efficient and scalable message routing, as the hash function distributes actors evenly across available shards, preventing load imbalances and reducing the likelihood of bottlenecks \cite{charousset2013native}.

\vspace{0.3cm}

By utilizing a hashing-based approach for actor-to-process mapping, the simulation can achieve high performance and scalability, effectively managing communication and resource allocation even as the number of actors increases substantially. This strategy optimizes computational resources, ensuring the system remains responsive and resilient under varying loads, which is crucial for accurately modeling complex and large-scale road traffic scenarios \cite{de_koster2016domains}.

\vspace{0.3cm}

In the geometric decomposition section, we discussed libraries that assist with graph partitioning for parallel programming. Some libraries help implement the actor model, but they have limitations. Libactor is one of them, catering to C programming and using threads for concurrency. However, it is not designed for distributed memory architectures. If we were to apply the Actor Model to the road simulation, we would need to use message-passing libraries \cite{haller2012actors}.

\vspace{0.3cm}

The most suitable library in this context would be MPI (Message Passing Interface). While MPI is not ideal for the Actor Model, as actors are supposed to create other actors dynamically for this simulation, which is challenging in MPI due to its fixed number of processes, it is still feasible. Additionally, while the Actor Model relies heavily on asynchronous, fire-and-forget messaging, and MPI does support non-blocking communication, it is not as intuitive or straightforward as in dedicated actor systems.

\vspace{0.3cm}

For the development of this program, C++ is recommended due to its support for object-oriented programming through classes, which effectively abstract the definitions of the Actor Model and provide a high-level representation of properties. This abstraction facilitates more accessible and efficient program development. In contrast, C relies on structs, which offer greater granular precision in defining components. Nevertheless, C++ affords sufficient granular control over the program while maintaining a higher level of abstraction than C. The advantages of C++'s abstraction mechanisms render the increased abstraction non-problematic, except in scenarios where maximum speed and efficiency are critical \cite{masud2018automatic}. Still, the rationale for the C++ recommendation is that we would need a base of abstraction in which we can build foundational classes rather than structs that interact more cohesively, as building these foundations with C structs would require significant nuance and take a lot of development time as we would have to create our ideologies from scratch. It would require more effort to build these programs; however, Rust may also be considered if expertise in the language is available, as it is a newer language with complex features that require careful programming. Other languages like Python and Go are impractical for programs that handle extensive simulations. Although it is possible to write simulations in these languages, the high levels of abstraction and the need for precise control would exacerbate efficiency deficiencies. Later, in this paper, we would discuss why high-level programming languages are not suitable for writing this road traffic simulation using a cellular automaton program as the subject of the experiment\cite{hewitt2015actor}.

\vspace{0.3cm}

High-Level Implementation Framework

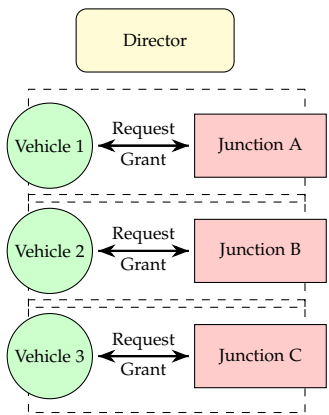
\begin{figure}[H]
\centering
\begin{tikzpicture}[
    % Removed scale or set to 1 for larger size
    scale=0.7,
    every node/.style={scale=0.7},
    vehicle/.style={
        circle, 
        draw, 
        minimum size=1.2cm, % Increased size
        fill=green!20,
        font=\normalsize
    },
    junction/.style={
        rectangle, 
        draw, 
        minimum width=2.5cm, % Increased width
        minimum height=1.2cm, % Increased height
        fill=red!20,
        font=\normalsize
    },
    summary/.style={
        rectangle, 
        draw, 
        rounded corners, 
        minimum width=3cm, 
        minimum height=1.2cm, 
        fill=yellow!20,
        font=\normalsize
    },
    message/.style={
        ->, 
        >=Stealth, 
        thick, % Increased line thickness
        shorten >=2pt, 
        shorten <=2pt
    },
    font=\normalsize % Set default font size
]

    % Actors
    \node[vehicle] (v1) at (0,0) {Vehicle 1};
    \node[vehicle] (v2) at (0,-2) {Vehicle 2};
    \node[vehicle] (v3) at (0,-4) {Vehicle 3};

    \node[junction] (jA) at (4,0) {Junction A};
    \node[junction] (jB) at (4,-2) {Junction B};
    \node[junction] (jC) at (4,-4) {Junction C};

    \node[summary] (summary) at (2,2) {Director};

    % Messages
    \draw[message] (v1) -- node[above, sloped] {Request} (jA);
    \draw[message] (v2) -- node[above, sloped] {Request} (jB);
    \draw[message] (v3) -- node[above, sloped] {Request} (jC);

    \draw[message] (jA) -- node[below, sloped] {Grant} (v1);
    \draw[message] (jB) -- node[below, sloped] {Grant} (v2);
    \draw[message] (jC) -- node[below, sloped] {Grant} (v3);

    % Background Dashed Boxes
    \begin{pgfonlayer}{background}
        % Increased inner sep for wider boxes
        \node[draw, dashed, fit=(v1) (jA), inner sep=0.5cm] {};
        \node[draw, dashed, fit=(v2) (jB), inner sep=0.5cm] {};
        \node[draw, dashed, fit=(v3) (jC), inner sep=0.5cm] {};
    \end{pgfonlayer}

\end{tikzpicture}
\caption{Mapping of Road Traffic Simulation Using the Actor Model}
\label{fig:actor-model-mapping}
\end{figure}

\paragraph{Conclusion}
It is important to note that the actor model is a powerful and precise model of computation, which makes it very important for the programmer to know what they are doing, as it is easy to produce subpar or incorrect code due to the complexity of implementing the Actor Model, which can make debugging and testing very time-consuming \cite{agha1997parallel}. Still, due to the problem's specifications, the Actor Model would be the most appropriate parallel strategy. The simulation is very dynamic, and numerous operations are happening at once. The future actions of the program are not predictable; it is not like a parallel heat equation solver or a cellular automaton problem, which would require writing the domain code and handling synchronization between processes; this problem is not as linear as those problems and would require significant complexity in synchronization. The reason why the Actor model is appropriate for this problem is that each actor manages itself independently and only acts on the messages it receives; we can program a message struct with numerous enums for each actor and decide the course of action based on the message enum and the variables within the struct \cite{latrous1995distributing}.

\vspace{0.3cm}

The Actor Model presents a robust and scalable parallel design strategy suitable for dynamic and complex simulations like road traffic modeling. Its ability to handle high concurrency, simplify synchronization, and maintain scalability makes it an ideal choice despite drawbacks such as message passing overhead and potential resource contention \cite{warwas2016actor}..

\subsection{Inappropriate Design Patterns}

\subsubsection{Pipeline}

\paragraph{Overview of the Pipeline Design Pattern}
Pipelining is a technique that enhances processing performance by dividing tasks into sequential stages, allowing multiple operations to overlap and execute concurrently. This procedure is used in various fields to improve efficiency and is prominently evident in computer science, particularly in parallel computing.

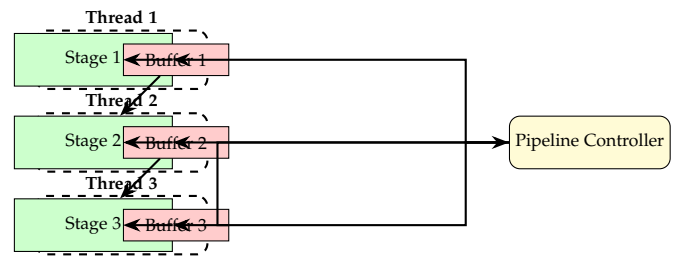
\begin{figure}[ht]
    \centering
    \begin{tikzpicture}[
        scale=0.55,
        every node/.style={scale=0.7},
        stage/.style={
            rectangle, 
            draw, 
            minimum width=3cm, 
            minimum height=1cm, 
            fill=green!20,
            font=\normalsize,
            align=center
        },
        sync/.style={
            rectangle, 
            draw, 
            minimum width=2cm, 
            minimum height=0.6cm, 
            fill=red!20,
            font=\normalsize,
            align=center
        },
        controller/.style={
            rectangle, 
            draw, 
            rounded corners, 
            minimum width=3cm, 
            minimum height=1cm, 
            fill=yellow!20,
            font=\normalsize,
            align=center
        },
        thread/.style={
            draw, 
            thick, 
            dashed, 
            rounded corners, 
            inner sep=0.2cm
        },
        arrow/.style={
            ->, 
            >=Stealth, 
            thick
        },
        font=\normalsize
    ]
        % Pipeline Controller
        \node[controller] (controller) at (8,0) {Pipeline Controller};
        
        % First define all stages and buffers
        \node[stage] (stage1) at (-4,2) {Stage 1};
        \node[sync] (stage1_sync) at (-2,2) {Buffer 1};
        
        \node[stage] (stage2) at (-4,0) {Stage 2};
        \node[sync] (stage2_sync) at (-2,0) {Buffer 2};
        
        \node[stage] (stage3) at (-4,-2) {Stage 3};
        \node[sync] (stage3_sync) at (-2,-2) {Buffer 3};
        
        % Then create the thread boxes
        \begin{pgfonlayer}{background}
            \node[thread, fit=(stage1)(stage1_sync), label=above:{\textbf{Thread 1}}] (thread1_box) {};
            \node[thread, fit=(stage2)(stage2_sync), label=above:{\textbf{Thread 2}}] (thread2_box) {};
            \node[thread, fit=(stage3)(stage3_sync), label=above:{\textbf{Thread 3}}] (thread3_box) {};
        \end{pgfonlayer}
        
        % Connections between stages and buffers
        \draw[arrow] (stage1) -- (stage1_sync);
        \draw[arrow] (stage1_sync) -- (stage2);
        \draw[arrow] (stage2) -- (stage2_sync);
        \draw[arrow] (stage2_sync) -- (stage3);
        \draw[arrow] (stage3) -- (stage3_sync);
        \draw[arrow] (stage3_sync) -- ++(1,0) |- (controller);
        
        % Control Signals from Controller to Stages
        \draw[arrow] (controller) -- ++(-3,0) |- (stage1);
        \draw[arrow] (controller) -- ++(-3,0) |- (stage2);
        \draw[arrow] (controller) -- ++(-3,0) |- (stage3);
    \end{tikzpicture}
    \caption{Pipeline Design Pattern in Parallel Computing Using Multiple Threads}
    \label{fig:pipeline-multi-thread}
\end{figure}

\paragraph{Unsuitability of Pipeline for Road Traffic Simulation}

A useful heuristic for determining whether the pipeline design strategy is appropriate is to assess if the overall computation involves performing calculations on multiple datasets that flow through a sequence of stages. This approach leverages concurrency in linear processes, similar to parallel computing \cite{navarro2009analytical}.

\vspace{0.3cm}

Applying this heuristic reveals that the current problem is not well-suited for the pipeline pattern. Although the issue is somewhat linear due to unidirectional graphs, dividing the computation into stages is impractical due to the dynamic nature of the simulation. While each junction could be represented as a pipeline stage, the computations at each intersection vary and depend on unpredictable sequences of events. The pipeline strategy is effective when there is a predetermined sequential order of operations at each step, which is not the case here \cite{mirsoleimani2018pipeline}. Consequently, it is unlikely to be an appropriate strategy for this or similar simulation problems. We discussed in the earlier sections that geometric decomposition is not suitable for not being able to predetermine computations effectively as simulations typically deal with randomness; we cannot account for randomness in the pipeline strategy as it is meant to follow a determined route \cite{chiu2022composing}.

\begin{figure}[h!]
\centering
\begin{tikzpicture}[
    scale=1,
    every node/.style={scale=0.6},  % Reduced from 0.8 to 0.6
    junction/.style={circle, draw=black, fill=gray!20, minimum size=0.8cm},
    vehicle/.style={rectangle, draw=black, fill=blue!20, minimum size=0.6cm},
    road/.style={rectangle, draw=black, fill=green!20, minimum size=0.6cm},
    >=Stealth
]
% Nodes
\node [vehicle] (v1) {Vehicle};
\node [road, right=1cm of v1] (r1) {Road};
\node [junction, right=1cm of r1] (j1) {Junction};
\node [road, right=1cm of j1] (r2) {Road};

% Arrows between components
\draw [->] (v1) -- node[above, font=\footnotesize] {Enter} (r1);  % Changed to \footnotesize
\draw [->] (r1) -- node[above, font=\footnotesize] {Travel} (j1);
\draw [->] (j1) -- node[above, font=\footnotesize] {Route Plan} (r2);

% Feedback loops
\draw [->, bend left=45] (j1.north) to node[above, font=\footnotesize] {Traffic Lights} (j1.north east);
\draw [->, bend left=45] (r1.south) to node[below, font=\footnotesize] {Congestion} (v1.south);

% Crash and fuel indicators
\node [below=0.7cm of j1, font=\footnotesize] (crash) {Possible Crash};
\draw [->, dashed, red] (j1) -- (crash);
\node [below=0.7cm of v1, font=\footnotesize] (fuel) {Fuel Depletion};
\draw [->, dashed, red] (v1) -- (fuel);

% Legend (more compact)
\node [vehicle, below=2cm of v1, xshift=1cm] (legend_vehicle) {};
\node [right=0.1cm of legend_vehicle, font=\footnotesize] {Vehicle};
\node [road, below=0.3cm of legend_vehicle] (legend_road) {};
\node [right=0.1cm of legend_road, font=\footnotesize] {Road};
\node [junction, below=0.3cm of legend_road] (legend_junction) {};
\node [right=0.1cm of legend_junction, font=\footnotesize] {Junction};
\end{tikzpicture}
\caption{\footnotesize Illustration of Complex Dependencies in the Road Simulation Model That Cannot Be Modeled Using a Pipeline}  % Caption also reduced
\label{fig:complex-dependencies}
\end{figure}
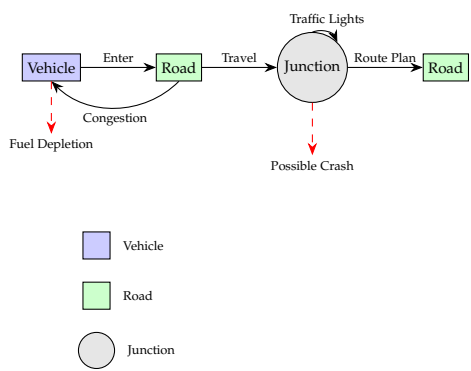

As shown in Figure~\ref{fig:complex-dependencies}, the road traffic simulation's complex dependencies demonstrate why a pipeline strategy would not be ideal for this problem.

\vspace{0.3cm}

Even if we somehow managed to implement the pipeline strategy, a significant question arises: how to map the pipeline stages effectively. Suppose we utilize OpenMP tasks and tasking features, even though pipelining isn't inherently supported. In that case, we could attempt to use OpenMP to manage the pipeline stages \cite{liu2013compiletime}. However, this raises the question: Should we treat each junction as a pipeline stage by assigning a separate thread to each stage? For a medium-sized simulation with 20,000 junctions, this approach would require 20,000 threads, which is impractical. The same issue arises when using processes, as the mapping would render this design strategy unsuitable for the problem \cite{preudhomme2012improvement}.

\vspace{0.3cm}

Synchronization between stages in the pipeline could eventually become a bottleneck. Another point is that pipelining is typically ideal for programs when broken into stages; each stage is equally computationally expensive or has an equal computational load. If one stage in the pipeline generally varies widely from the median, the slowest stage would become a bottleneck. In the case of the road traffic simulation, we are not aware beforehand of the computation distribution of the problem, hindering the ability to load balance effectively \cite{bell1991pipelined}.

\subsubsection{Recursive Data}

\paragraph{Overview of the Recursive Data Design Pattern}

The recursive data design strategy is a programming approach where data structures are defined in terms of themselves. This self-referential definition allows for the creation of complex, hierarchical data models like linked lists, trees, and nested objects. Using recursion, developers can simplify handling data that naturally fits into nested or hierarchical patterns, making code more intuitive and easier to manage for specific problems \cite{nawaz2011recursive}.

\begin{figure}[ht]
    \centering
    \begin{tikzpicture}[
      scale=1.0,  % Added overall scale reduction
      level distance=1.2cm,  % Reduced from 1.5cm
      level 1/.style={sibling distance=4cm},  % Reduced from 6cm
      level 2/.style={sibling distance=2cm},  % Reduced from 3cm
      level 3/.style={sibling distance=1cm},  % Reduced from 1.5cm
      every node/.style={
          circle, 
          draw, 
          fill=green!20, 
          minimum size=0.8cm,  % Reduced from 1cm
          align=center,
          font=\footnotesize  % Added smaller font
      },
      edge from parent/.style={
          draw, 
          -{Stealth[length=2mm, width=1.5mm]},  % Reduced arrow size
          thick, 
          color=purple
      },
    ]
    
    % Recursive Binary Tree Structure
    \node {Root}
        child {
            node {Left}
            child { 
                node {L-L}
                child { node {LL-L} }
                child { node {LL-R} }
            }
            child { 
                node {L-R}
                child { node {LR-L} }
                child { node {LR-R} }
            }
        }
        child {
            node {Right}
            child { 
                node {R-L}
                child { node {RL-L} }
                child { node {RL-R} }
            }
            child { 
                node {R-R}
                child { node {RR-L} }
                child { node {RR-R} }
            }
        };
    
    \end{tikzpicture}
    \caption{\footnotesize Example of Recursive Data on an Extended Binary Tree}  % Reduced caption font
    \label{fig:binary-tree-recursive}
\end{figure}
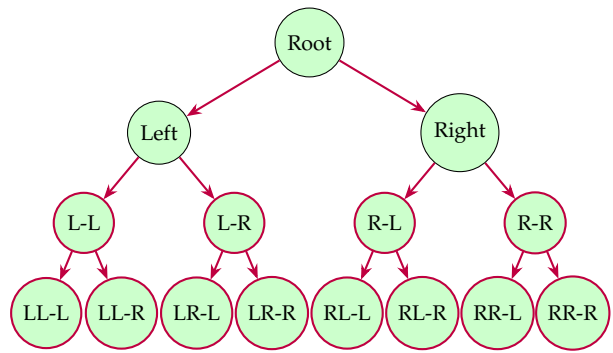

\vspace{0.3cm}

A Recursive Data pattern operates on a sequential data structure. As Figure~\ref{fig:binary-tree-recursive} highlights, it would be appropriate for programs requiring sequential processing. It can also be used for graph processing, but the main reason it isn't suitable for the current simulation problem is the aspect of the simulation dynamics.

\vspace{0.3cm}

\paragraph{Unsuitability of Recursive Data for Road Traffic Simulation}

Understanding that the recursive data problem involves a distributed recursive function is essential. Problems that can be solved recursively on a single processing element (PE) could, in theory, be adapted to a distributed recursive data pattern. However, when considering the specific problem it cannot be solved recursively. Additionally, writing a recursive program to solve a simulation problem is currently not feasible.

\vspace{0.3cm}

Implementing the recursive strategy in a serial context could involve a depth-first search algorithm on a tree, where the program reaches its base case and then pops off elements from a stack to combine the results.

\vspace{0.3cm}

In a parallel context, it could imply a parallel depth-first or breadth-first search in a graph where parallelism occurs at each node and its neighbors, allowing multiple edges to be processed simultaneously. This strategy would be excellent for standardized graph problems with deterministic computations at each node, as it will enable the time complexity to be broken down from an $O(N)$ algorithm to an $O(\log N)$ algorithm. However, this is not the case for the current problem because, similar to why the pipeline strategy is unsuitable, we cannot predict what computations will be done at each stage. We do not know the calculations that would be done at each edge of the graph, as different simulation parameters can lead to computations that cannot be accounted for beforehand \cite{nishimura1999parallel, ahn2007analysis, misra1994powerlist}.

\vspace{0.3cm}

Another rationale is why recursive data is not a suitable approach. The simulation represents a network of roads and junctions modeled as a graph, where roads are unidirectional edges and junctions are nodes. Vehicles move through this network, and their interactions depend on dynamic factors like traffic lights, congestion, and fuel levels. The road network's simulation dynamics do not naturally lend themselves to recursive data modeling because it involves complex, non-hierarchical relationships and cycles \cite{hwang1998interprocedural, benczur2004static, westbrook1989algorithms}.

\vspace{0.3cm}

Using recursive data structures to model the road network could lead to complications such as circular references, which are challenging to manage and cause issues like infinite loops or stack overflows \cite{corbera2000accurate, arnold1998parallel}. Additionally, the simulation requires efficient algorithms for route planning, congestion management, and vehicle movement, all of which benefit from iterative processing and data structures optimized for graph traversal, such as adjacency lists or matrices \cite{hendren1989parallelizing, navarro2012data}.

\vspace{0.3cm}

Moreover, the simulation involves real-time updates to the state of vehicles, roads, and junctions. Vehicles need to recalculate routes based on current road speeds and traffic light statuses, and the system must handle events like crashes and fuel depletion. Implementing these dynamic behaviors using recursive data design would introduce unnecessary complexity and potential performance bottlenecks \cite{bianchini2002recursive, klarlund1993graph}. Iterative algorithms are better suited for handling the mutable state and frequent updates required in the simulation

\vspace{0.3cm}

The rationale for choosing the Actor model stems from the significant synchronization required in this program, which would be unsuitable for linear computation models. Recursive data patterns would struggle to handle this level of synchronization, and even if we attempted to use such a pattern, structuring the graph to integrate it effectively would be highly impractical \cite{hwang1997identifying, chu2017local}. Integrating this pattern within this framework is likely more impractical than using the pipeline design pattern.

\subsection{Approaches to Evaluate Parallel Code}

To evaluate the proposed parallel program implementing the actor model for this road simulation using MPI, we would need to employ a combination of performance metrics, analytical methods, and specialized tools to thoroughly assess its efficiency and scalability \cite{hpc_perf_eval}. Key metrics such as running time, speedup, parallel efficiency, and convergence properties provide valuable insights into how well the parallelization performs relative to the original serial code \cite{kumar1994introduction}; good experimental practice also has to take place to ensure the reliability and consistency of findings \cite{jain1991art}. Additionally, understanding the optimal mapping of actors to processors, knowing how to segment the actors efficiently that allows clean code and scalability, and determining the optimal processor count is also crucial for maximizing performance \cite{actor_perf}.

\vspace{0.3cm}

\paragraph{Measuring Running Time and Speedup}
Firstly, measure the \textbf{running time} $T_p$ of the parallel program on $p$ processors. By comparing this to the running time of the serial program $T_1$, compute the \textbf{speedup} $S_p$, defined as \cite{amdahl1967validity}:
\[
S_p = \frac{T_1}{T_p}
\]
This metric would help indicate how much faster the parallel program runs than the serial version. Ideally, we should aim for linear speedup, where $S_p = p$, but in practice, the speedup is often sub-linear due to overheads such as communication and synchronization and other factors that are not accounted for \cite{gustafson1988reevaluating}.

\vspace{0.3cm}

\paragraph{Calculating Parallel Efficiency}
We should also calculate the \textbf{parallel efficiency} $E_p$ using \cite{eager1989speedup}:
\[
E_p = \frac{S_p}{p} = \frac{T_1}{p \cdot T_p}
\]
Parallel efficiency measures how effectively the processors are utilized in the parallel program. A high efficiency close to $1$ signifies that the processors are being used optimally, whereas a low efficiency indicates overheads diminishing the benefits of parallelization \cite{karp1990optimal}. Analyzing $E_p$ assesses the program's scalability and identifies diminishing returns as the processor count increases; this would help identify the optimal processor count and also unveil insights into our program that may not be obvious.

\vspace{0.3cm}

\paragraph{Utilizing Performance Analysis Tools}
We also need to thoroughly evaluate these metrics and utilize various \textbf{performance analysis tools} designed explicitly for parallel programs \cite{nagel1996vampir}. Numerous profiling tools can provide insights about the bottlenecks in our parallel program; there is an MPI profiling tool called Scalasca \cite{geimer2010scalasca}, which can pinpoint hotspots and areas for optimization within our code. We can use this to understand tradeoffs in our implementation details, such as using persistent communication rather than point-to-point communication. These tools can profile MPI applications to identify bottlenecks in communication and computation \cite{knupfer2008vampir}. They collect data on MPI function calls, message sizes, and frequencies, helping understand the overhead introduced by inter-process communication.

\vspace{0.3cm}

\paragraph{Optimizing Actor-to-Processor Mapping}
We can also examine the \textbf{optimum mapping of actors to processors} \cite{vernon1988performance}. In the actor model implemented with MPI, actors represent entities such as vehicles, roads, or junctions, and their mapping to processors affects communication patterns and load balancing \cite{buyya1999high}. Experiment with different mapping strategies, such as:

\vspace{0.3cm}

\begin{itemize}
\item \textbf{Geographic Partitioning}: Assigning actors based on their spatial locality in the simulation to minimize communication between processors \cite{pellegrini1996scotch}.
\item \textbf{Load Balancing Algorithms}: Using dynamic load balancing techniques to distribute actors evenly across processors to prevent some processors from becoming bottlenecks \cite{devine2002zoltan}.
\end{itemize}

\vspace{0.3cm}

\paragraph{Assessing Convergence Properties}
The most essential thing in every computational process is the result. If the result is not accurate, then the whole program is not useful, which is why we would need \textbf{convergence properties} to ensure that parallelization does not adversely affect the accuracy or stability of the simulation results \cite{dongarra2003numerical}. Verify that the parallel program produces results consistent with the serial version by comparing key outputs \cite{bailey1991parallel}.

\vspace{0.3cm}

\paragraph{Conducting Scalability Testing}
Another crucial component is to conduct \textbf{scalability testing}. Perform both \textbf{strong scaling} and \textbf{weak scaling} tests \cite{hager2010introduction}. This would help evaluate the parallel program's performance under different scaling scenarios and provide us with insights that other evaluations couldn't.

\vspace{0.3cm}

\vspace{0.3cm}
All these tools should be used strategically, as they complement each other and don't work in isolation; they could be used interchangeably. It should be known when to use a tool to evaluate and conclude the result of another tool, as these would provide the most insights and intuitions about the program.

\section{Conclusion}
Parallel computing is a central aspect of modern computer science applied to anything from mundane smartphone to supercomputer usage. Following its path of evolution back to the very origins—theoretical foundations of PRAM and BSP to modern multi-core and heterogeneous systems—it is possible to appreciate both the principles that are the foundations of parallelism and the practical considerations needed to parallelize successfully \cite{valiant1990bridging}.

\vspace{0.3cm}
Central to this is the understanding that parallel performance is a function of several interrelated factors. The intrinsic character of a problem, notably the ease with which a problem can be decomposed into parts that can then be distributed among tasks, usually determines the degree to which parallelization can provide meaningful speedups \cite{gustafson1988reevaluating}. Additionally, hardware elements like CPUs, caches, memory, and special-purpose accelerators add to the complications since load balancing, cache coherence, and memory hierarchy directly impact the efficiency of execution \cite{hennessy2011computer}. No less significant are the software models of programming—processes, threads, message passing, actor-oriented designs, and a broad range of parallel patterns (such as geometric decomposition, pipeline, and recursive data)—each contributing models of coordinating work among distributed concurrent environments \cite{mcCool2012structured}.

\vspace{0.3cm}
Beyond foundational principles, this effort also strongly emphasizes quality tooling and a supportive environment. Profiling and debugging software like Intel VTune, Allinea DDT, or NVIDIA Nsight, combined with well-engineered programming frameworks (MPI, OpenMP and CUDA), supply the capabilities to develop, refine, and support scalable parallel applications \cite{shafi2009multi}. Concurrent with this are shared knowledge bases and information that supply the means to learn constantly and improve together. As the domain pushes ahead—confronting challenges of exascale computing, parallelism in the cloud, and AI workloads—these parallelization skills will increasingly become paramount to master. Understanding problem decompositions, hardware-software interactions, and contemporary debugging and optimizing techniques is not merely advantageous to the domain of HPC but also echoes all areas of the computing domain. Computer science can tackle the concurrent future with strong, scalable parallel programs by incorporating historical knowledge, theoretic models, and practical applications into their designs \cite{dongarra2011international}.

\newpage
\bibliographystyle{IEEEtran}
% Generated by IEEEtran.bst, version: 1.14 (2015/08/26)

\newpage
\section{Appendix}

\subsection{Details of the Cirrus supercomputer}

The Cirrus HPC system is located at the Uni- versity of Edinburgh. The system comprises 283 compute nodes, each with 2.1 GHz, 18-core Intel Xeon E5-2695 processors and 256 GB of memory.
The Cirrus supercomputer features 36 GPU compute nodes. Each node is equipped with two Intel Xeon Gold 6148 (Cascade Lake) processors running at 2.5 GHz, with each processor containing 20 cores. These cores support 2 hardware threads (Hyperthreads) per core, which are enabled by default. Addi- tionally, each node contains four NVIDIA Tesla V100-SXM2-16GB (Volta) GPU accelerators, which are interconnected and connected to the host processors via PCIe. In total, the GPU compute nodes provide 144 GPU accelerators (4 GPUs × 36 nodes) and 1,440 CPU cores (40 cores × 36 nodes) across the system. The cache hierarchies for the compute nodes are detailed in Table 3

\begin{table}[h]
\centering
\caption{Cache Hierarchies for CPU and GPU Nodes}
\label{tab:cache}
\begin{minipage}{0.45\textwidth}
\centering
\subcaption{CPU Node Cache Hierarchy}
\begin{tabular}{ll}
\hline
\textbf{Cache Level} & \textbf{Size} \\
\hline
L1 Cache & 32 KiB per core \\
L2 Cache & 256 KiB per core \\
L3 Cache & 45 MiB shared \\
\hline
\end{tabular}
\end{minipage}%
\hfill
\begin{minipage}{0.45\textwidth}
\centering
\subcaption{GPU Node Cache Hierarchy}
\begin{tabular}{ll}
\hline
\textbf{Cache Level} & \textbf{Size} \\
\hline
L1 Cache & 32 KiB per core \\
L2 Cache & 1 MiB per core \\
L3 Cache & 27.5 MiB shared \\
\hline
\end{tabular}
\end{minipage}
\end{table}
\subsection{Pseudocode for MPI-Based Cellular Automaton Simulation}

\begin{algorithm}[H]
\caption{Main Simulation Procedure}
\begin{algorithmic}[1]
\Procedure{Main}{}
    \State $args \gets$ \Call{ParseArguments}{}
    \If{$args.mode = \texttt{"serial"} \lor$ MPI\_Size() $= 1$}
        \State \Call{RunSimulationSerial}{$args$}
    \Else
        \State \Call{RunSimulationParallel}{$args$}
    \EndIf
\EndProcedure
\end{algorithmic}
\end{algorithm}

\bigskip

\noindent \textbf{Procedure: RunSimulationSerial($args$)}
\begin{algorithmic}[1]
    \State Set parameters: $L, \rho, seed, maxstep, printfreq \gets args$
    \State Initialize random generator with $seed$
    \State $grid \gets$ ZeroMatrix$(L+2, L+2)$ \Comment{Create grid with ghost boundaries}
    \For{each cell $(i,j)$ in $grid[1:L,1:L]$}
        \State Set $grid[i,j] \gets 1$ with probability $\rho$, else $0$
    \EndFor
    \State $initial\_live \gets$ Sum$(grid[1:L,1:L])$
    \State \textbf{Print} simulation parameters
    \State Start timer
    \For{$step = 1$ to $maxstep$}
        \State \Call{UpdateGhostBoundaries}{$grid$} \Comment{Enforce periodic conditions}
        \State \Call{ApplyBoundaryConditions}{$grid, L$}
        \State $neighbors \gets$ \Call{ComputeNeighbors}{$grid$} \Comment{Sum of 4-adjacent cells}
        \State $live \gets$ \Call{UpdateCells}{$grid, neighbors$} \Comment{Cells become alive if neighbor count is in \{2,4,5\}}
        \If{$step \mod printfreq = 0$}
            \State \textbf{Print} current step and live cell count
        \EndIf
        \If{$live < 0.75 \times initial\_live \lor live > 1.33 \times initial\_live$}
            \State \textbf{Print} termination message and \textbf{break}
        \EndIf
    \EndFor
    \State Stop timer and compute average time per iteration
    \State Write $grid[1:L,1:L]$ to output file
\end{algorithmic}

\bigskip

\noindent \textbf{Procedure: RunSimulationParallel($args$)}
\begin{algorithmic}[1]
    \State Set parameters: $L, \rho, seed, maxstep, printfreq \gets args$
    \State Initialize MPI and create Cartesian topology: obtain $comm$, $rank$, $size$, $dims$, $coords$, $cart\_comm$
    \State Compute local dimensions: $(local\_rows, local\_cols) \gets L/dims$
    \If{$rank = 0$}
        \State $global\_grid \gets$ CreateRandomMatrix$(L, L)$ with probability $\rho$
    \EndIf
    \State $global\_grid \gets$ Broadcast$(global\_grid, \text{root}=0)$
    \State $local\_grid \gets$ ExtractSubgrid$(global\_grid, coords, local\_rows, local\_cols)$
    \State $grid \gets$ CreateExtendedGrid$(local\_grid, local\_rows, local\_cols)$
    \State $local\_initial\_live \gets$ Sum$(local\_grid)$
    \State $initial\_live \gets$ ReduceSum$(local\_initial\_live)$ and Broadcast to all processes
    \If{$rank = 0$}
        \State \textbf{Print} simulation parameters and number of processes
    \EndIf
    \State Barrier synchronization on $cart\_comm$
    \State Start timer (MPI\_Wtime)
    \For{$step = 1$ to $maxstep$}
        \State \Call{ExchangeHalos}{$grid, cart\_comm, local\_rows, local\_cols$}
        \State \Call{AdjustBoundaries}{$grid, coords, dims, local\_rows, local\_cols, L$}
        \State $neighbors \gets$ \Call{ComputeNeighbors}{$grid$}
        \State $local\_live \gets$ \Call{UpdateCells}{$grid, neighbors$}
        \State $total\_live \gets$ ReduceSum$(local\_live)$ over all processes
        \If{$rank = 0 \land (step \mod printfreq = 0)$}
            \State \textbf{Print} current step and $total\_live$
        \EndIf
        \If{$total\_live < 0.75 \times initial\_live \lor total\_live > 1.33 \times initial\_live$}
            \If{$rank = 0$}
                \State \textbf{Print} termination message and \textbf{break}
            \EndIf
        \EndIf
    \EndFor
    \State Stop timer and compute average time per iteration (on $rank=0$)
    \State $global\_result \gets$ GatherSubgrids$(grid[1:local\_rows, 1:local\_cols])$
    \If{$rank = 0$}
        \State Write $global\_result$ to output file
    \EndIf
    \State Finalize MPI
\end{algorithmic}

\subsection{Details of the Road Simulation Model}

\vspace{0.4cm}
\begin{itemize}
    \item The road map is represented as a graph:
    \begin{itemize}
        \item Junctions are nodes in the graph, and roads are edges between these nodes.
        \item Individual roads (graph edges) are always unidirectional (one way), so always connect one junction (graph node) to another and not the other way round.
        \item Whilst there will be routes connecting a specific junction to many others in the graph, there is no guarantee that there is a route between every pair of junctions.
        \item The graph is provided to the model via an input file, which is read during initialisation.
    \end{itemize}
    
    \item Roads are all different lengths, expressed in metres:
    \begin{itemize}
        \item Roads also have a maximum vehicle speed and current speed. The current speed depends upon how congested the road is (i.e., the number of vehicles currently on the road).
        \item The road’s length and maximum vehicle speed are also specified in the initialisation file.
    \end{itemize}

    \item Vehicles travel from one junction to another:
    \begin{itemize}
        \item Vehicles are periodically added to the simulation, and at this point, their source (starting junction) and destination junction are set.
        \item While the source and destination junctions are random, only those with valid routes between them will be selected.
    \end{itemize}

    \item Junctions may have multiple roads connecting them to other junctions:
    \begin{itemize}
        \item A junction may or may not have traffic lights.
        \item If traffic lights are present:
        \begin{itemize}
            \item Only one road connected to the junction can be used at any time.
            \item Vehicles requiring other roads must wait until the traffic lights enable their desired road.
            \item Traffic lights change each simulated minute, working in a round-robin manner to enable one road after another.
        \end{itemize}
        \item If no traffic lights are present, all roads connected to the junction can be used at all times.
        \item Traffic light information is included in the initialisation file.
    \end{itemize}

    \item Vehicles travel on roads and through junctions:
    \begin{itemize}
        \item There are six types of vehicles: car, bus, mini-bus, coach, motorbike, and bike.
        \begin{itemize}
            \item Each has different capabilities, including maximum speed and maximum passenger capacity.
            \item These parameters are defined in the code.
            \item When a vehicle is created, the number of passengers is randomly generated up to the maximum capacity.
        \end{itemize}
        \item While on a road, vehicles continually update their location based on their current speed.
        \begin{itemize}
            \item The speed at which a vehicle travels depends on the road's speed when the vehicle left the junction and the vehicle’s maximum speed (whichever is smaller).
            \item Vehicles do not continuously check the road’s current speed after entering it.
        \end{itemize}
        \item When a vehicle arrives at a junction:
        \begin{itemize}
            \item If the junction is not its destination, the vehicle re-plans its route.
            \item Route planning considers the road lengths and, for most roads, their maximum speed.
            \item For roads connected to the current junction, the vehicle uses the current speed instead of the maximum speed.
            \item If the junction has traffic lights, the vehicle waits until the selected road is enabled.
            \item If the junction is the vehicle’s destination, it is removed from the simulation.
        \end{itemize}
    \end{itemize}

    \item Vehicle crashes can occur at junctions without traffic lights:
    \begin{itemize}
        \item All crashes involve a single vehicle (i.e., no multi-vehicle collisions).
        \item Crashes only occur at junctions, never on roads.
        \item The likelihood of a crash increases with the number of vehicles at the junction.
        \item A crashed vehicle is removed from the simulation.
    \end{itemize}

    \item Vehicles have a finite amount of fuel, which can run out:
    \begin{itemize}
        \item When a vehicle is created, its fuel amount (expressed in seconds of running time) is randomly generated.
        \item Fuel consumption is continuous, whether the vehicle is moving or waiting at a junction.
        \item Vehicle speed does not affect fuel consumption.
        \item When fuel runs out, the vehicle is removed from the simulation.
    \end{itemize}

    \item The simulation periodically prints progress summaries to \texttt{stdio}, including:
    \begin{itemize}
        \item Current simulation time (in minutes).
        \item Total number of vehicles added to the simulation.
        \item Total number of passengers delivered to their destination.
        \item Total number of passengers stranded (due to crashes or fuel depletion).
        \item Number of vehicle crashes.
        \item Number of vehicles that have run out of fuel.
    \end{itemize}

    \item The simulation terminates after a predetermined number of simulation minutes:
    \begin{itemize}
        \item A final summary is printed to \texttt{stdio}, including:
        \begin{itemize}
            \item Current simulation time (in minutes).
            \item Total number of vehicles added to the simulation.
            \item Total number of passengers delivered to their destination.
            \item Total number of passengers stranded.
            \item Number of vehicle crashes.
            \item Number of vehicles that ran out of fuel.
        \end{itemize}
        \item A file is written with more detailed statistics:
        \begin{itemize}
            \item For each junction, the number of crashes and the number of vehicles that have passed through.
            \item For each road, the number of vehicles that traveled on it and the highest concurrent vehicle count (i.e., congestion levels).
        \end{itemize}
    \end{itemize}
\end{itemize}

\end{document}